\documentclass[fleqn,usenatbib]{mnras}

\usepackage{newtxtext,newtxmath}

\usepackage[T1]{fontenc}
\usepackage{ae,aecompl}

\usepackage{graphicx}	
\usepackage{amsmath}	
\usepackage{amssymb}	

\title[Galaxy formation in the P-Millennium]{Galaxy formation in the Planck Millennium: the atomic hydrogen content of dark matter halos}

\author[C. M. Baugh et al.]
{\parbox{18cm}{
C. M. Baugh,$^{1}$\thanks{E-mail: c.m.baugh@durham.ac.uk (CMB)}
Violeta Gonzalez-Perez$^{2}$,
Claudia D.P. Lagos$^{3}$,
Cedric G. Lacey$^1$,
John C. Helly$^1$,
Adrian Jenkins$^1$,
Carlos S. Frenk$^1$,
Andrew J. Benson$^4$,
Richard G. Bower$^1$,
Shaun Cole$^1$
}
\\
\\
$^{1}$Institute for Computational Cosmology, Department of Physics, Science Laboratories, Durham University, \\
South Road, Durham, DH1 3LE, UK\\
$^{2}$Institute of Cosmology and Gravitation, Portsmouth University, Dennis Sciama Building, Burnaby Road, Portsmouth PO1 3FX, UK\\
$^{3}$International Centre for Radio Astronomy Research, 7 Fairway, Crawley, 6009, Perth, WA, Australia\\
$^{4}$Carnegie Observatories, 813 Santa Barbara Street, Pasadena, CA 91101, USA\\
}

\date{Accepted XXX. Received YYY; in original form ZZZ}

\pubyear{2018}

\begin{document}
\label{firstpage}
\pagerange{\pageref{firstpage}--\pageref{lastpage}}
\maketitle

\begin{abstract}
We present recalibrations of the {\tt GALFORM} semi-analytical model of galaxy formation in a new N-body simulation with the Planck cosmology. The Planck Millennium simulation uses more than 128 billion particles to resolve the matter distribution in a cube of $800$ Mpc on a side, which contains more than 77 million dark matter haloes with mass greater than $2.12 \times 10^{9} h^{-1} {\rm M_{\odot}}$ at the present day. Only minor changes to a very small number of model parameters are required in the recalibration. We present predictions for the atomic hydrogen content (HI) of dark matter halos, which is a key input into the calculation of the HI intensity mapping signal expected from the large-scale structure of the Universe. We find that the HI mass $-$ halo mass relation displays a clear break at the halo mass above which AGN heating suppresses gas cooling, $\approx 3 \times 10^{11} h^{-1} M_{\rm \odot}$. Below this halo mass, the HI content of haloes is dominated by the central galaxy; above this mass it is the combined HI content of satellites that prevails. We find that the HI mass - halo mass relation changes little with redshift up to $z=3$. The bias of HI sources shows a scale dependence that gets more pronounced with increasing redshift.  
\end{abstract}




\section{Introduction}

Measuring fluctuations in the intensity of $21\,$cm line emission offers a novel way to map the large-scale structure of the Universe that is competitive with the largest planned optical galaxy redshift surveys \citep{Bull:2015}. The $21\,$cm line is a forbidden transition between hyperfine structure in the ground state of  atomic hydrogen. As a consequence, redshift surveys of galaxies detected through their weak HI emission at best currently contain thousands rather than the hundreds of thousands or even millions of galaxies reached by optically selected surveys (for a review of extragalactic HI astronomy see \citealt{Giovanelli:2015}). The next  generation of surveys, such as the Widefield ASKAP L-band Legacy All-sky Blind surveY, WALLABY, the Australian Square Kilometer Array Pathfinder, will measure HI for over half a million galaxies \citep{Johnston:2008,Duffy:2012}. However, such HI surveys will still be limited to the local Universe. 
A solution to this problem is to exploit the finite angular and frequency resolution of radio telescopes, to effectively stack the emission from many galaxies in a single pointing and hence boost the HI signal to a measurable level. Measuring the intensity of HI emission from all the sources within some volume bypasses the challenge of detecting the emission from individual sources, allowing a  view of the large-scale structure of the Universe to be obtained that is smoothed on small scales \citep{Battye:2004,Zaldarriaga:2004, Peterson:2006,Pritchard:2012, KOvetz:2017}.

The power spectrum of HI intensity fluctuations has already been measured in spite of the cosmological signal being much smaller than the galactic foreground \citep{Switzer:2013}. The fact that this measurement was made at a much higher redshift ($z \sim 0.8$) than that for which estimates of the HI mass function  are available by the measurement of emission from single galaxies ($z \sim 0.05$) illustrates the potential of intensity mapping. Encouraged by this, a number of HI intensity mapping experiments are either under construction or proposed (e.g. BAO from Integrated Neutral Gas Observations (BINGO) \citealt{Battye:2016}; the Canadian Hydrogen Intensity Mapping Experiment (CHIME) Pathfinder \citealt{Bandura:2014}, the Five-hundred-meter Aperture Spherical radio Telescope (FAST) \citealt{Bigot-Sazy:2016}, MeerKAT \citealt{Pourtsidou:2017} and the Square Kilometre Array (SKA) \citealt{Santos:2015}).

A key input into the prediction for HI intensity fluctuations is the HI content of dark matter halos, combining the contribution from the central galaxy with that of all of the satellite galaxies in the halo. Many empirical approaches have been proposed to describe the HI content of dark matter haloes  including: 1) simple scalings with halo mass \citep{Santos:2015}, 2) arguments based on the effective circular velocity of halos that contain HI, limited at low circular velocities by photo-ionization heating of the intergalactic medium and at high velocities as a result of the central galaxy tending, with increasing halo mass, to become  bulge rather than disk dominated, and hence gas poor (\citealt{Barnes:2009, Bagla:2010}), 3) more sophisticated scalings with halo mass, with a broken power law and variable amplitude, constrained to fit various observations of the abundance and clustering of HI galaxies \citep{Padmanabhan:2015,Pad:2016,Padmanabhan:2017a,Padmanabhan:2017b,Obuljen:2018,Paul:2018}. Several studies using hydrodynamical simulations have yielded predictions for the HI galaxy mass $-$ halo mass relation by post-processing the simulation results to divide the cold gas mass of a galaxy into atomic and molecular components\footnote{By `atomic' and `molecular' gas we mean phases of the interstellar medium in which the hydrogen is preferentially in atomic (HI) or molecular (H$_{2}$) form, respectively.}  \citep{Dave:2013,Villa:2014,Crain:2017,Villaescusa:2018,Ando:2018}.  

Here we use a different approach to predict the form of the HI mass $-$ halo mass relation and its evolution: a physically motivated model of galaxy formation in which the atomic and molecular gas contents of model galaxies are tracked at all times. Semi-analytical models calculate the transfer of baryons between different reservoirs within dark matter halos that are growing hierarchically \citep{Baugh:2006, Benson:2010,SomervilleDave2015}. Comparisons between the predictions made by models developed by different groups, which are publicly available, reveal that the models have reached a level of maturity such that they can give robust predictions for the baryonic content of dark matter haloes and the clustering of samples defined by different properties, such as stellar mass, cold gas mass and star formation rate\footnote{This is true for samples defined by galaxy number density, after ranking the galaxies by the value of a property such as stellar mass or cold gas mass. The models differ in the observations used to set their parameters, and so some distribution functions will agree between models (such as the stellar mass function) better than others (such as the cold gas mass function).}  \citep{Contreras:2013,Guo:2016,Pujol:2017,Lagos:2018}.  \cite{Contreras:2015} examined the dependence of galaxy properties on the mass of the host halo (or the mass of the host subhalo at infall for the case of satellites). This study revealed that some properties, such as stellar mass, display a simple dependence on host halo mass, albeit with considerable scatter (see also the review by \citealt{Wechsler:2018}). Other properties, however, such as cold gas mass and star formation rate, are predicted to have a complicated dependence on halo mass. Hence we cannot simply translate trends uncovered by the analysis of stellar mass selected samples and assume that these hold for HI-selected  samples; a physical model is needed to connect the HI mass of a galaxy to its host halo mass.   

Here, we use the {\tt GALFORM} semi-analytical galaxy formation model \citep{Cole:2000, Baugh:2005, Bower:2006,GonzalezPerez:2014,Lacey:2016}. The treatment of star formation in {\tt GALFORM} was extended by \cite{Lagos:2011} to model the atomic and molecular hydrogen  contents of galaxies, rather than just the total cold gas mass that was considered in earlier versions of the model (note that some other semi-analytical codes now also have this capability: \citealt{Fu:2010,Somerville:2015,Stevens:2017,Xie:2017,Lagos:2018}). A comprehensive overview of {\tt GALFORM} and the way in which the model predictions respond to varying the galaxy formation parameters can be found in \cite{Lacey:2016}. 

\cite{Kim:2015} used the {\tt GALFORM} model of \cite{Lagos:2012} to investigate the physics behind the low mass end of the HI mass function, arguing that the photoionization heating of the intergalactic medium was the key process shaping this prediction. The same model was used by \cite{Kim:2017} to predict the HI content of halos and the HI intensity fluctuation power spectrum. Here we update the background cosmological model used in {\tt GALFORM} with a new N-body simulation, the Planck Millennium. We recalibrate the versions of {\tt GALFORM} calibrated using halo merger trees extracted from an N-body simulation run with the WMAP-7 cosmological parameters \citep{Guo:2013}. In particular, we consider the models introduced by \cite{Lacey:2016} and  \cite{GonzalezPerez:2014} (see also \citealt{Guo:2016,GP:2018}). We recalibrate the galaxy formation parameters which define these models to reflect the change in the cosmological parameters, the improved mass and time resolution of the Planck Millennium simulation outputs and the incorporation of an improved treatment of galaxy mergers (\citealt{Simha:2017}, see also \citealt{Campbell:2015}). We then use the recalibrated models to predict the HI content of dark matter halos and its evolution. The recalibrated models described here have been used by \cite{Cowley:2018} to make predictions for the counts and redshift distributions of galaxies that will be seen by the James Webb Space Telescope and by \cite{Nuala:2017} to test the assumptions behind halo occupation distribution modelling of galaxy clustering. The Planck Millennium simulation has also been used by \cite{Veena:2018} to study the spin and shape alignments of haloes in the cosmic web.

The layout of this paper is as follows. In Section 2, we introduce the galaxy formation model, describing the Planck Millennium N-body simulation (\S~2.1), giving an outline of {\tt GALFORM} (\S~2.2), explaining the differences and similarities of the two variants of {\tt GALFORM} considered (\S~2.3) and closing with the recalibration of these models (\S~2.4).  The predictions for the HI content of dark matter halos are given in \S~3, along with comparisons to previous results, and our conclusions are given in \S~4. Some other predictions relating to intensity mapping using tracers other than HI are presented in Appendix~\ref{sec:IM}, and a variant of the recalibrated models with gradual ram pressure stripping of the hot gas in satellite galaxy halos is discussed in Appendix~\ref{sec:ram}. Note that throughout we quote masses in units of $h^{-1} \, {\rm M_{\odot}}$ and lengths in units of $h^{-1} {\rm Mpc}$, retaining the reduced Hubble parameter, $h$, where $H_{0} = 100 \, h \, {\rm km} \, {\rm s}^{-1} \,{\rm Mpc}^{-1}$.

\section{The galaxy formation model}

We first (\S~\ref{sec:pmill}) introduce the Planck Millennium N-body simulation, putting it in the context of the other simulations in the Millennium suite. In \S~\ref{sec:galform} we give a brief overview of the {\tt GALFORM} model. We  explain the similarities and differences between the variants introduced by \cite{GonzalezPerez:2014} and \cite{Lacey:2016} in~\S2.3. The recalibration of the parameters defining these models for their implementation in the Planck Millennium is presented in~\S2.4.

\subsection{The P-Millennium N-body simulation}
\label{sec:pmill} 

The Planck Millennium N-body simulation (hereafter the PMILL) is the latest in the `Millennium' series of simulations of structure formation in the dark matter in cosmologically representative volumes carried out by the Virgo Consortium (see Table~\ref{table:nbody} for a summary of the specifications of these runs and the cosmological parameters used). The PMILL follows the evolution of the matter distribution in a similar but slightly larger volume (by a factor of $\times \, 1.43$, after taking into account the differences in the Hubble parameters assumed, see Table~\ref{table:nbody}) than the simulation  described by \cite{Guo:2013} (hereafter WM7). The cosmological parameters used in the PMILL correspond to those of the best-fitting basic six parameter cold dark matter model for the first year Planck cosmic microwave background data and measurements of the large-scale structure in the galaxy distribution (\citealt{Planck}); these parameter values have changed little with the analysis of the final Planck dataset \citep{Planck:final:cosmo}. The cosmological parameters used in the PMILL are exactly the same as those used in the {\tt EAGLE} simulations of \cite{Schaye:2015}.

The PMILL uses over 128 billion particles ($5040^{3}$) to represent the matter distribution, which is more than an order of magnitude more than was used in the MSI or WM7 runs. This, along with the simulation volume used, places the PMILL at an intermediate resolution between the MSI of \cite{Springel:2005} and the MSII run described in \cite{Boylan-Kolchin:2009}.

The initial conditions were generated at $z=127$  using second order Lagrangian perturbation theory as set out in \cite{Jenkins:2010}. The random 
phases used to generate the initial conditions are taken from the public {\it Panphasia} Gaussian white noise field \citep{Jenkins:2013} and correspond to those of the 800 Mpc {\tt EAGLE} simulation volume listed in Table B1 of \cite{Schaye:2015}.
The simulation was run on 4096 processors of the COSMA-4 machine of the DiRAC installation at Durham, using a reduced memory version of the N-body code {\tt GADGET} \citep{GADGET}, and took around $20$ Tb of RAM. The halo and subhalo finder {\tt SUBFIND} \citep{Springel:2001} was run concurrently with the simulation, accounting for around a quarter of the {\tt CPU} time; in total, the simulation and halo finding took 7 million {\tt CPU} hours. A single full particle output of the simulation is around $3.8$ Tb. The halo data is around $0.5$ Tb per output, depending on the redshift. The PMILL run has many more outputs than the MSI, with the halos and subhalos stored at 271 redshifts compared with the $~60$ outputs used in the MSI. Dark matter halo merger trees were constructed from the {\tt SUBFIND} subhalos using the {\tt DHALOS} algorithm described in \cite{Lilian:2014} (see also \citealt{Merson:2013}). Halos are retained that contain at least 20 particles, corresponding to a halo mass resolution limit of $2.12 \times 10^{9} \, h^{-1} {\rm M_{\odot}}$. The full particle data is only stored at selected snapshots; the full particle and halo data for all 271 outputs would correspond to  a dataset of size $1 $Pb. 

\begin{figure}
    	\includegraphics[clip, trim=0.cm 1.2cm 0cm 0.5cm,width=1.05\columnwidth]{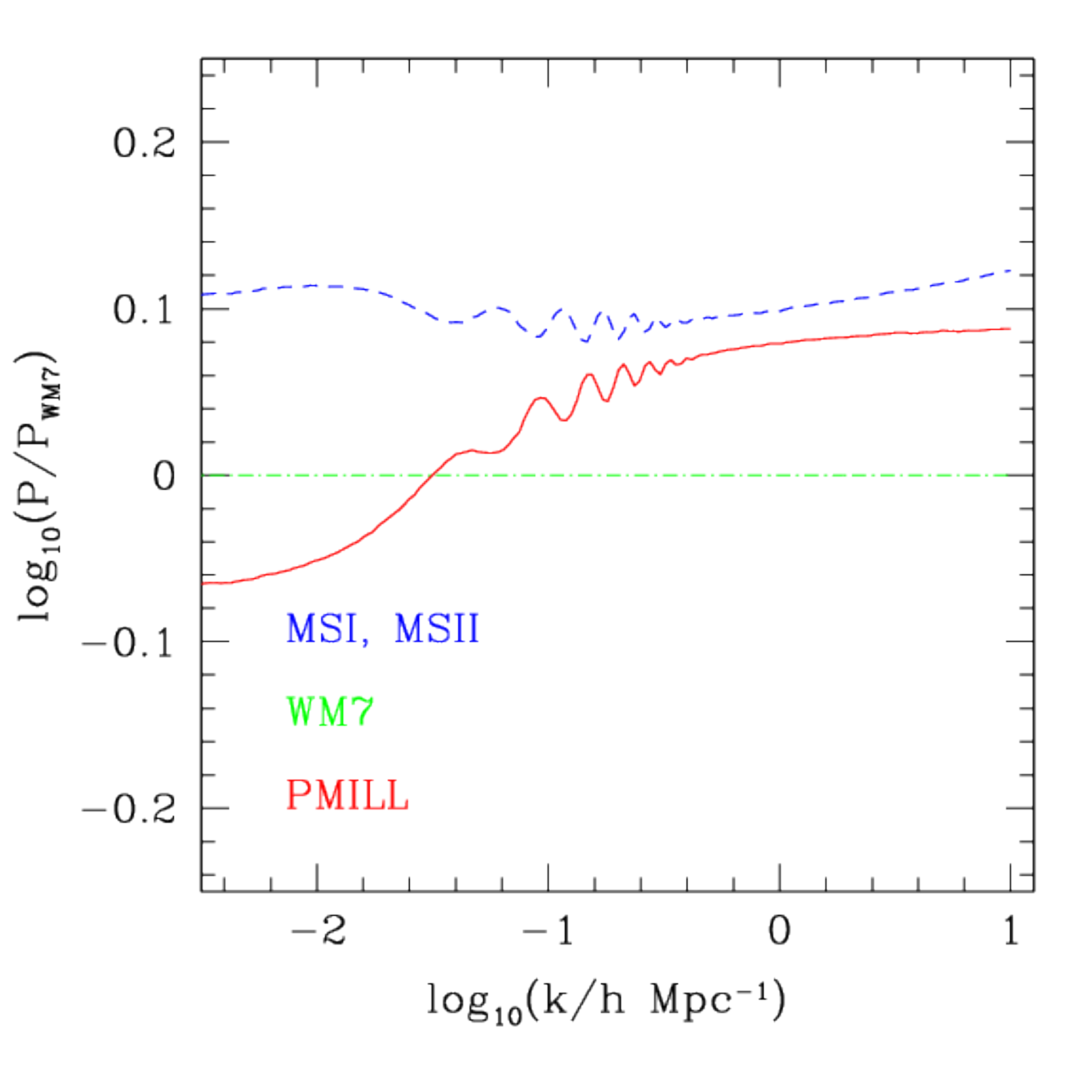}
    \caption{The linear perturbation theory power spectra used in the Millennium Simulations, plotted as ratios to the power spectrum used in the WM7. The labels refer to the simulation names listed in Table~1.
    }
    \label{fig:pk}
\end{figure}

\begin{table*} 
	\caption{Selected parameters of the Millennium N-body simulations.The first column gives the present day matter density in units of the critical density (note that in all cases, the cosmology used corresponds to a flat universe, with the remainder of the critical energy density made up by a cosmological constant), column (2) gives the baryon density parameter, (3) the spectral index of the primordial density fluctuations, (4) the reduced Hubble parameter, $h = H_{0}/( 100 \, {\rm km \, s^{-1}} \, {\rm Mpc}^{-1})$, (5) the normalisation of the density fluctuations at the present 
    day, (6) the simulation box length, (7) the number of particles, (8) the particle mass, and (9) the halo mass limit corresponding to  20 particles. Column (10) gives the label used to refer to the simulation in the text and column (10) gives the reference for the simulation.}
	\begin{tabular}{cccccccccll} 
		\hline
		$\Omega_{\rm M}$ & $\Omega_{\rm b}$ & $n_{\rm spec}$ & $h$ & $\sigma_{8}$ & $L_{\rm box}$        & $N_{\rm p}$           & $M_{\rm p}$ & $ M_{\rm h}$ & Label & Reference \\
		                 &                  &            &           &              &$(h^{-1} \, {\rm Mpc})$  &     & $(h^{-1}\, {\rm M_{\rm \odot}})$ & $( h^{-1} \, {\rm M_{\odot}})$ &        &         \\
		\hline
		0.25\hphantom{0} & 0.0455\hphantom{0} & 1.0\hphantom{000} & 0.73\hphantom{00} & 0.9\hphantom{000} & 500\hphantom{000} 
  & $ 2160^{3} $ & $8.61 \times 10^{8}$ & $1.72 \times 10^{10}$& MSI & \cite{Springel:2005}\\
		0.25\hphantom{0} & 0.0455\hphantom{0} & 1.0\hphantom{000} & 0.73\hphantom{00} & 0.9\hphantom{000} & 100\hphantom{000}  & $ 2160^{3} $ & $6.88 \times 10^{6}$ & $1.37 \times 10^{8}$& MSII & \cite{Boylan-Kolchin:2009}\\
		0.272 & 0.0455\hphantom{0} & 0.967\hphantom{0} & 0.704\hphantom{0} & 0.81\hphantom{00} & 500\hphantom{000}  & $ 2160^{3} $ & $9.34 \times 10^{8}$ & $1.87 \times 10^{10}$& WM7 & \cite{Guo:2013}\\
		0.307 & 0.04825 & 0.9611 & 0.6777 & 0.8288 & 542.16  & $ 5040^{3} $ & $1.06 \times 10^{8}$ & $2.12 \times 10^{9}$& PMILL & This paper\\
		\hline
	\end{tabular}
    \label{table:nbody}
\end{table*}

Table~\ref{table:nbody} lists the cosmological parameters used in the Millennium simulations. The linear perturbation theory power spectra corresponding to these cosmological models are compared in Fig.~\ref{fig:pk}. The power spectrum of density fluctuations used in the PMILL run is similar to that used in the MSI and MSII runs on the scales that are most relevant to the growth and abundance of dark matter halos ($k > 0.1 \, h \, {\rm Mpc}^{-1}$), with an amplitude a little under 10\% higher than 
the spectrum used in the WM7 run. On large scales (small wavenumbers), the PMILL power spectrum has less power than the MSI power spectrum due to the higher value of $\Omega_{\rm M} h$, which means that matter-radiation equality happens sooner in the PMILL cosmology than in the cosmology used in the Millennium, and because of the tilt in the spectral index in the PMILL cosmology. Hence, the length scale of the turnover in the power spectrum is smaller in the PMILL cosmology (which means it occurs at a higher wavenumber). 

The halo mass functions in the PMILL and WM7 are very close to one another. For halo masses $> 1 \times 10^{12} \, h^{-1} \, {\rm M_{\odot}}$, the PMILL mass function is $\approx 0.1$ dex higher in halo abundance at a given mass than the WM7 one. This difference could be removed by rescaling the halo masses in the PMILL down by 10\, \%. The physical density of baryons, $\propto \Omega_{\rm b} h^{2}$, agrees to within $1.5\, \%$ in the two simulation cosmologies, which implies that there will be little difference in the gas cooling rates in halos of similar mass in the PMILL and WM7 cosmologies.

\subsection{The GALFORM semi-analytical galaxy formation model}
\label{sec:galform}

{\tt GALFORM} is used to make an {\it ab initio} calculation of the formation and evolution of galaxies during the hierarchical growth of structure in the dark matter (\citealt{Cole:2000,Bower:2006,GonzalezPerez:2014,Lacey:2016}; for reviews of hierarchical galaxy formation, see \citealt{Baugh:2006}, \citealt{Benson:2010} and \citealt{Somerville:2015}). {\tt GALFORM} models the following processes: (i) the formation and merging of DM halos; (ii) the shock-heating and radiative cooling of gas inside DM halos, leading to the formation of galactic disks; (iii) star formation (SF) in galaxy disks and in starbursts; (iv) feedback from supernovae (SNe), from AGN and from photo-ionization of the IGM;
(v) galaxy mergers driven by dynamical friction and bar instabilities in galaxy disks, both of which can trigger starbursts and lead to the formation of spheroids;
(vi) calculation of the sizes of disks and spheroids; (vii) chemical enrichment of stars and gas. The reprocessing of starlight by dust, leading to both dust extinction at UV to near-IR wavelengths, and dust emission at far-IR to sub-mm
wavelengths, is calculated self-consistently from the gas and metal contents of each galaxy and the predicted scale lengths of the disk and bulge components using a radiative transfer model (see \citealt{Lacey:2016} for a description of the implementation of dust extinction in {\tt GALFORM}).

A thorough description of how each of these processes is modeled in {\tt GALFORM} is set out in \cite{Lacey:2016}. Here, for completeness, we recap the description of selected processes for which the parameters are varied later on to recalibrate the models for the PMILL simulation. This section could be skipped by the reader who is pressed for time. 

\subsubsection{Supernova feedback}
Supernovae inject energy into the ISM, which causes gas to be ejected from galaxies. The rate of gas ejection due to supernova feedback is assumed to be proportional to the instantaneous star formation rate, { $\psi$}, with a mass loading factor $\beta$ that is taken to be a power law in the galaxy circular velocity $V_{\rm c}$:
\begin{equation}
{\dot M}_{\rm eject} = \beta(V_{\rm c}) \, \psi = \left( \frac{V_{\rm c}}{V_{\rm SN}} 
\right)^{-\gamma_{\rm SN}} \, \psi .
\label{eq:Meject}
\end{equation}
The circular velocity used is that at the half-mass radius of the disk for disk star formation, and of the spheroid for starbursts.  This formulation uses two adjustable parameters: $\gamma_{\rm SN}$, which specifies the dependence of $\beta$ on circular velocity, and $V_{\rm SN}$ which gives the normalization. We assume that cold gas is ejected from a galaxy to beyond the virial radius of its host dark matter halo.

Gas ejected from the galaxy in this way by SN feedback is assumed to accumulate in a reservoir of mass $M_{\rm res}$ beyond the virial radius, from where it gradually returns to the hot gas reservoir within the virial radius, at a rate
\begin{equation}
{\dot M}_{\rm return} = \alpha_{\rm ret} 
\frac{M_{\rm res}}{\tau_{\rm dyn,halo}} ,
\label{eq:Mreturn}
\end{equation} 
where $\tau_{\rm dyn,halo} = r_{\rm vir}/V_{\rm vir}$ is the halo dynamical time and $\alpha_{\rm ret}$ is a parameter. 

\subsubsection{AGN feedback}
Supermassive black holes (SMBHs) release energy through accretion of gas, making them visible as AGN, and leading to a further  feedback process on galaxy formation. In {\tt GALFORM}, SMBHs grow in three ways \citep{Malbon:2007,Bower:2006,Fanidakis:2011,Griffin:2018}: (i) accretion of gas during starbursts triggered by galaxy mergers or disk instabilities ({\em starburst mode}); (ii) accretion of gas from the hot halo ({\em hot halo mode}); (iii) BH-BH mergers. The mass accreted onto the SMBH in a starburst is assumed to be a constant fraction, $f_{\rm BH}$, of the mass which is turned into stars, where $f_{\rm BH}$ is a parameter. We assume that AGN feedback occurs in the {\em radio mode} \citep{Croton:2006,Bower:2006}: energy released by direct accretion of hot gas from the halo onto the SMBH powers relativistic jets that deposit thermal energy in the hot halo gas which can balance energy losses from  radiative cooling. This radio-mode feedback is assumed to set up a steady state in which the energy released by the SMBH accretion exactly balances the radiative cooling, if both of the following conditions are met: (a) the cooling time of halo gas is sufficiently long compared to the free-fall time
\begin{equation}
\tau_{\rm cool}(r_{\rm cool})/\tau_{\rm ff}(r_{\rm cool}) > 1/\alpha_{\rm cool} ,
\label{eq:alpha_cool}
\end{equation}
where $\alpha_{\rm cool} \sim 1$ is an adjustable parameter (with larger values resulting in more galaxies being affected by AGN feedback); and (b)
the AGN power required to balance the radiative cooling luminosity
$L_{\rm cool}$ is below a fraction $f_{\rm Edd}$ of the Eddington luminosity $L_{\rm Edd}$ of the SMBH of mass $M_{\rm BH}$
\begin{equation}
L_{\rm cool} < f_{\rm Edd} L_{\rm Edd}(M_{\rm BH}) .
\label{eq:f_Edd}
\end{equation}

\subsubsection{Star formation in disks}
The star formation rate (SFR) in galactic disks is calculated using the empirical law derived from observations of nearby star-forming disk galaxies by \cite{Blitz:2006}, as implemented in {\tt GALFORM} by \cite{Lagos:2011}. The cold gas in the disk is divided into atomic and molecular phases, with the local ratio of surface densities
$\Sigma_{\rm atom}$ and $\Sigma_{\rm mol}$ at each radius in the disk depending on the gas pressure, $P$, in the midplane through 
\begin{equation}
R_{\rm mol} = \frac{\Sigma_{\rm mol}}{\Sigma_{\rm atom}} 
= \left( \frac{P}{P_0} \right)^{\alpha_{\rm P}} .
\label{eq:at_mol}
\end{equation}
We use $\alpha_{\rm P}=0.8$ and $P_0/k_{\rm B} = 17\,000~{\rm cm}^{-3} {\rm K}$ based on observations \citep{Leroy:2008}. The pressure is calculated from the surface densities of gas and stars, as described in \citet{Lagos:2011}. The SFR is then assumed to be proportional to the mass in the molecular component only; integrated over the whole disk this gives a star formation rate 
\begin{equation}
\psi_{\rm disk} =  \nu_{\rm SF} M_{\rm mol,disk} = 2 \pi \int_{0}^{\infty} \nu_{\rm SF} \Sigma_{\rm mol} \, r \, {\rm d r} ,
\label{eq:SFR_mol}
\end{equation}
where $f_{\rm mol} = R_{\rm mol}/(1+R_{\rm mol})$ and $M_{\rm mol} $ is the mass of molecular gas in the disk. \citet{Bigiel:2011} find a best-fitting value of $\nu_{\rm SF} = 0.43 \, {\rm Gyr}^{-1}$ for a sample of local
galaxies, with a $1\sigma$ range of 0.24~dex. The disk SFR law (\ref{eq:SFR_mol}) has a non-linear dependence on the total cold gas mass through the
dependence on $f_{\rm mol}$.

\subsubsection{Photoionsation heating feedback}

Ionizing photons produced by stars and AGN ionize and heat the IGM, restricting   galaxy formation in two ways: (i) the increased IGM pressure inhibits the collapse  of  gas  into  dark  matter  halos;  (ii)  photo-heating of  gas  inside  halos  by  the  ionizing  UV  background  inhibits  the cooling  of  gas.  These  effects  are modelled by  assuming  that  after the IGM is reionized at a redshift $z= z_{\rm reion}$, no further cooling of gas occurs  in  halos  with  circular  velocities $V_{\rm vir} < V_{\rm crit}$.  We
adopt $z_{\rm reion} = 10$ (e.g  \citealt{Dunkley:2009}), and
$V_{\rm crit} = 30 \,{\rm km \,s}^{-1}$ as suggested by gas  dynamical  simulations \citep{Okamoto:2008}. \cite{Kim:2015} demonstrate how varying $z_{\rm reion}$ and $V_{\rm crit}$ changes the model predictions for the low mass end of the HI mass function.

\subsubsection{Galaxy mergers}

In the WM7 versions of the {\tt GALFORM} models considered here, the timescale for the merging of satellites with the central galaxy in their host halo due to dynamical friction is computed following the general method  described in \cite{Cole:2000}. This approach assumes that when a new halo forms, each satellite galaxy enters the halo on a random orbit. A merger timescale is then computed using an analytical formula. While the \cite{GonzalezPerez:2014} model makes use of the equations presented by \citet{Lacey:1993}, a modified expression for the merger timescale is used in the model of \cite{Lacey:2016}. The latter expression has been fitted to numerical simulations to account for the tidal stripping of subhaloes \citep{Jiang:2008}, but otherwise the treatment is the same; i.e. an analytic merger time-scale is computed as soon as a galaxy enters a larger halo. The satellite is considered to have merged with its central galaxy once the merger time-scale has elapsed, provided that this transpires before the halo merges to form a larger system, in which case a new merger time-scale is computed. Note that this scheme does not take into account that the satellite galaxy may still be associated with a resolvable dark matter subhalo at the time the galaxy merger takes place. Tests of the model predictions for galaxy clustering on small scales \citep{Contreras:2013} and the radial distribution of galaxies in clusters \citep{Budzynski:2012} indicate that this merger scheme results in satellite galaxy  distributions that are too centrally concentrated.

In the recalibration of the models presented here, we instead use an improved treatment of galaxy mergers that was developed by \cite{Simha:2017} and was first used in {\tt GALFORM} by \cite{Campbell:2015}. \citet{GP:2018} used this merger scheme in a model calibrated in the WM7 simulation. The new scheme is more faithful to the subhalo information from the underlying N-body simulation,  reducing the reliance on analytically determined orbits and timescales. Satellite galaxies track the positions of their associated subhaloes. When the subhalo hosting a satellite can no longer be resolved following mass stripping, the position and velocity of the subhalo when it was last identified are used to compute an analytical  merger timescale. This timescale is then used in the same way as in the default scheme described above. The merger timescale calculation assumes a \cite{Navarro:1997} halo mass distribution to compute the orbital parameters of the satellite at the time its subhalo was last identified, combined with a modified version of the analytical time-scale used by \cite{Lacey:1993}. If a halo formation event occurs at a time after the subhalo is lost, a new merger time-scale for the satellite is calculated in the same way, using instead the position and velocity of the particle which was the most bound particle of the subhalo when it was last identified. In the improved {\tt GALFORM} merger scheme, a satellite galaxy is not considered as a candidate for merging while it remains associated with a resolved subhalo.

\subsection{The starting points: the Gonzalez-Perez et~al. and Lacey et~al. variants of {\tt GALFORM}}

\begin{figure}
	\includegraphics[trim={0.8cm 1cm 1.cm  0.4cm},clip,width=1.02\columnwidth]{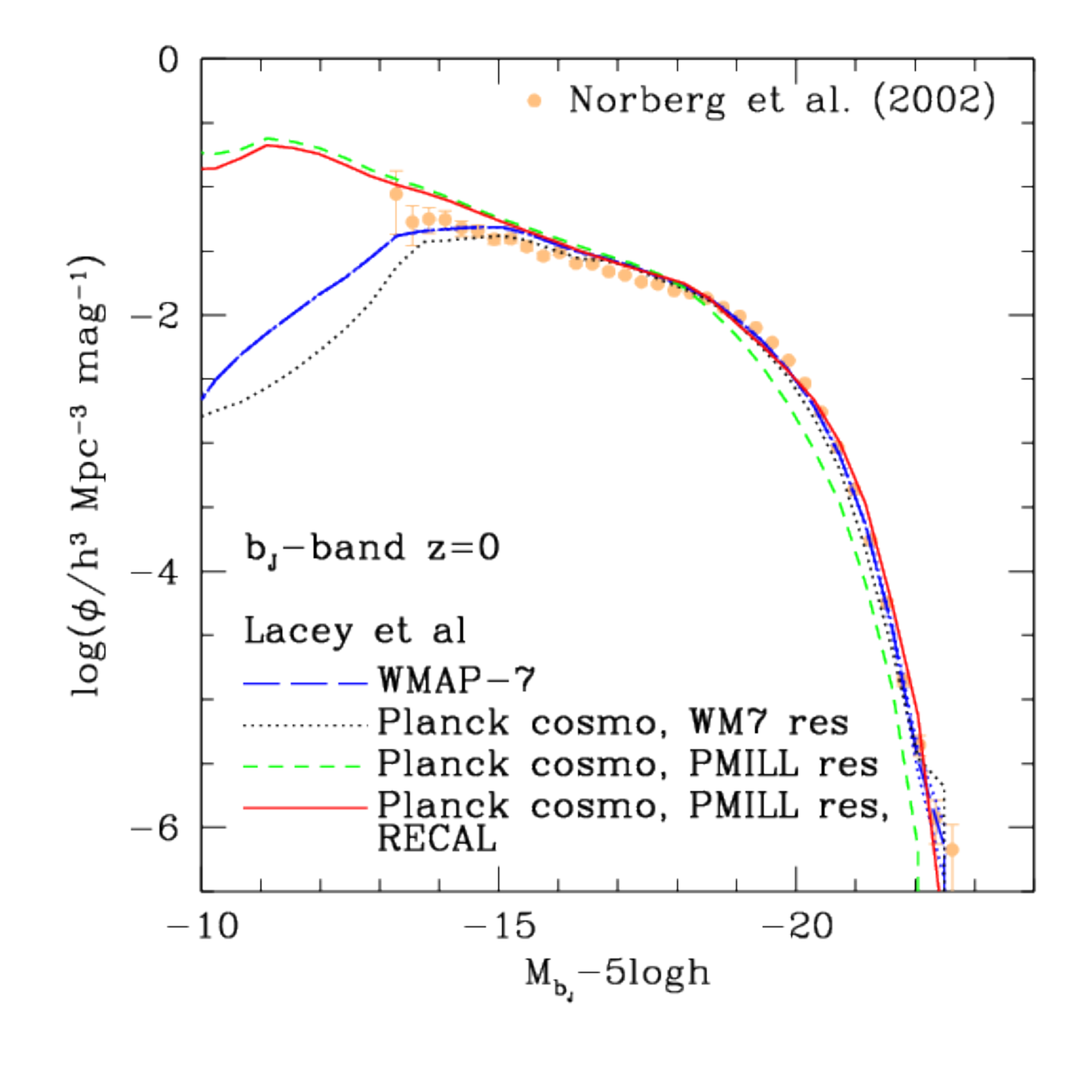}
    	\includegraphics[trim={0.8cmcm 1.5cm 1.cm 0},clip,width=1.02\columnwidth]{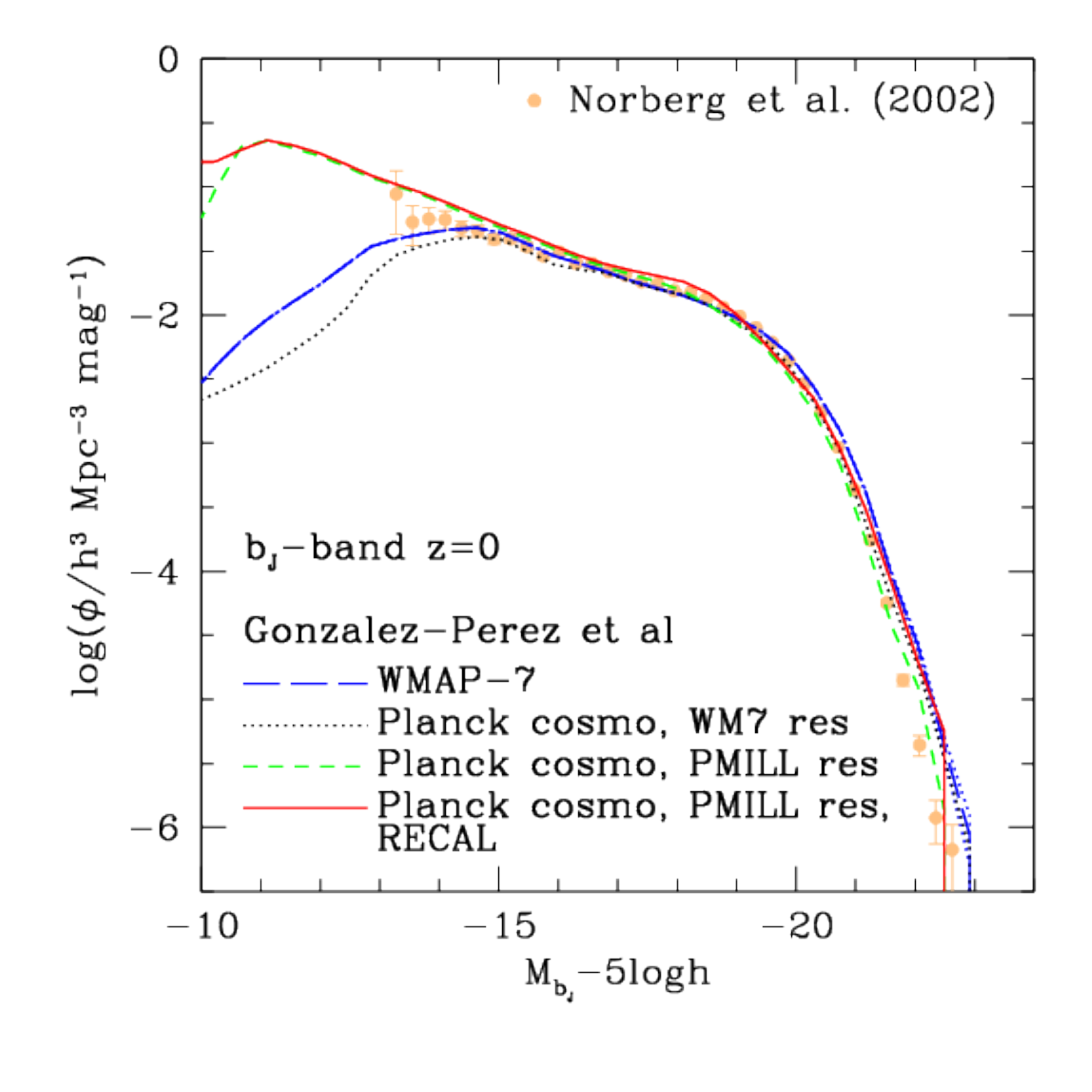}
    \caption{The $b_{\rm J}$-band galaxy luminosity function at $z=0$, which is one of the primary datasets used to calibrate the {\tt GALFORM} model parameters. The symbols show the observational estimate from \citet{Norberg:2002}. The lines show various {\tt GALFORM} predictions, as labeled by the legends, based on the \citet{Lacey:2016} (top) and \citet{GonzalezPerez:2014} (bottom) models. The blue dashed lines show the fiducial models in the WM7 run. The black dotted lines show the luminosity function obtained with the same galaxy formation parameters and merger tree resolution as used in the WM7 run, but changing the cosmology to that of the PMILL. The green dashed line shows the same case, but now using the full resolution of the PMILL trees. The red line shows the recalibrated version of each model for the PMILL.}
    \label{fig:lf}
\end{figure}

\begin{figure}
	\includegraphics[trim={0.8cm 1cm 0  0.5cm},clip,width=1.05\columnwidth]{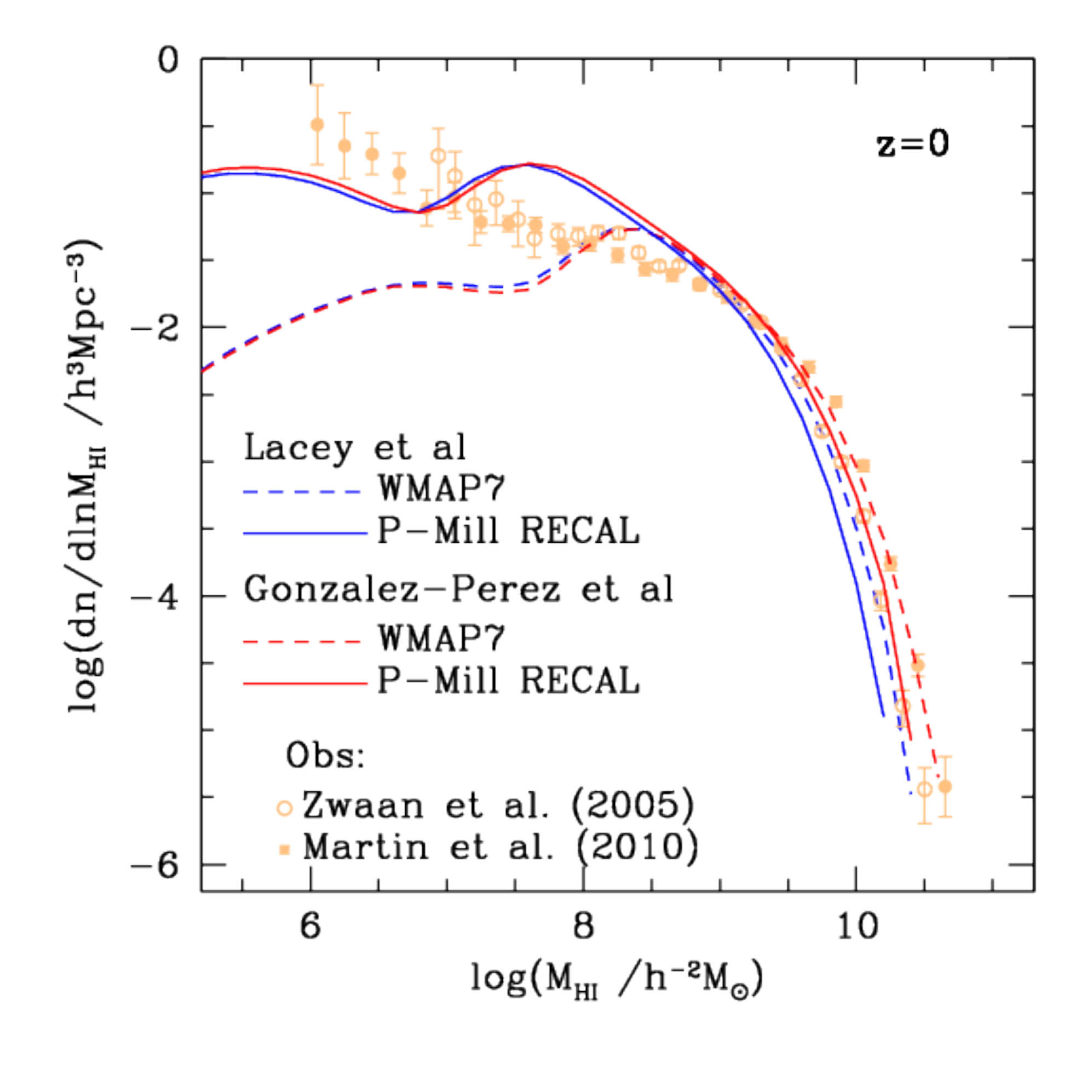}
    	\includegraphics[trim={0.8cm 1cm 0 0},clip,width=1.05\columnwidth]{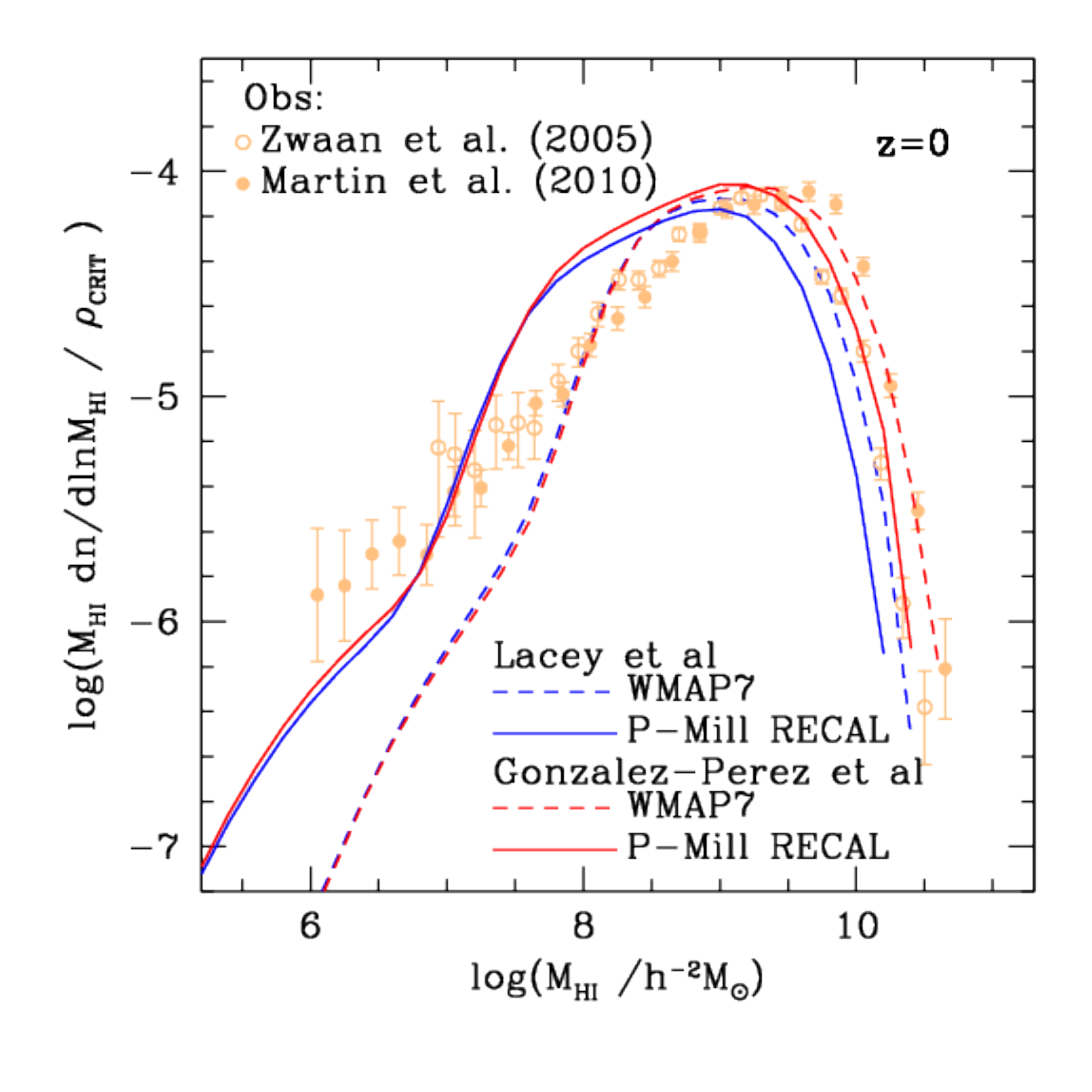}

    \caption{{\it Upper panel:} The HI mass function of galaxies at $z=0$. The symbols show observational estimates as labelled. The dashed lines show the predictions of the \citet{Lacey:2016} (blue) and \citet{GonzalezPerez:2014} (red) models in the WM7 run. The solid lines show the predictions of the recalibrated versions of these models in the PMILL. {\it Lower panel:} Same as the upper panel, but now the contribution of each galaxy is weighted by its HI mass.}
    \label{fig:himf}
\end{figure}

The parameters of the \cite{GonzalezPerez:2014} and \cite{Lacey:2016} {\tt GALFORM} models were calibrated using the dark matter halo merger histories from the WM7 N-body simulation. The model parameters were chosen to reproduce the local galaxy luminosity function in the $b_{\rm J}$ and $K$-bands (see Fig.~\ref{fig:lf}, discussed later), along with a range of other, mostly local datasets, as discussed in \cite{Lacey:2016}. Both models show reasonable agreement with the luminosity function of galaxies at high redshift in the rest-frame $UV$ and $K$-bands. In addition, the \cite{Lacey:2016} model was designed to match the number counts of galaxies detected by their emission at long wavelengths ($250 - 850 \mu {\rm m}$). To achieve this, \cite{Lacey:2016} invoked a mildly top-heavy initial mass function (IMF) for stars made in bursts resulting from dynamically unstable disks or galaxy mergers. Hence the primary difference between the two models is the choice of IMF in star formation that takes place in bursts. Both models assume a solar neighbourhood IMF for quiescent star formation in disks. \citet{GonzalezPerez:2014} also assume this IMF for stars produced in bursts. There are slight differences in the values chosen for some of the other galaxy formation parameters as a result of the assumptions about the IMF.

\subsection{Recalibration for the P-Millennium N-body simulation}

The first step in our recalibration is to see what the \cite{GonzalezPerez:2014} and \cite{Lacey:2016} model predictions look like when changing the cosmological parameters from those used in WM7 to the ones used in the PMILL run whilst keeping all of the other parameters fixed. For this exercise we prune the halo merger trees extracted from the PMILL to use a resolution equivalent to that in the WM7 simulation (after taking into account the difference in $\Omega_{\rm M}$, this corresponds to imposing a minimum halo mass of $2.11 \times 10^{10} \, h^{-1} \, {\rm M_{\odot}}$ in the merger trees). Fig.~\ref{fig:lf} shows that the present-day galaxy luminosity function hardly changes on adopting the PMILL cosmological parameters (comparing the blue dashed and black dotted lines). This is to be expected given the minor changes in the matter power spectrum, halo mass function and physical density of baryons on changing the cosmological parameters, as discussed in~\S2.1. 

If we now use the full resolution of the PMILL halo merger trees, i.e. retaining halos down to the 20 particle limit of $2.11 \times 10^{9} h^{-1} {\rm M_{\rm \odot}}$, Fig.~\ref{fig:lf} shows that the model luminosity function retains a power law form at the faint end down to four magnitudes fainter than in the WM7 case, corresponding to a factor of 40 in luminosity. Using the full resolution of the halo merger trees, the models now predict many more faint, low stellar mass galaxies. There is also a small reduction in the number of bright galaxies on improving the resolution of the halo merger trees. This is model dependent, with a more noticeable change in the Lacey et~al. model than with the Gonzalez-Perez et~al. parameters. This small model dependence of the predictions to changing the mass resolution of the merger trees is consistent with the results of  \cite{Guo:2011}, who found essentially no difference in the predictions of the {\tt L-GALAXIES} model for the abundance of bright galaxies comparing the outputs of the MSI and MSII simulations; the merger trees for these simulations differ in mass resolution by a factor of 125 (see Table~1). 

Our aim is to make minimum number of changes to the parameters necessary to recalibrate the Lacey et~al. and Gonzalez et~al. models for use with the PMILL merger trees. We therefore use the same calibration data, mainly the present-day optical and near-infrared galaxy luminosity functions used to set the parameters in the original models, and do not attempt to improve upon the level of agreement shown by the predictions of the original models. We also take this opportunity to use the improved treatment of galaxy mergers proposed by \cite{Simha:2017}. The results of the recalibration are shown by the red lines in Fig.~\ref{fig:lf}. We found it necessary to make minor adjustments to just two parameters to obtain the recalibrations for the PMILL: 1) for Lacey et~al. RECAL, we changed $\gamma_{\rm SN}$ (Eqn.~\ref{eq:Meject}) from $3.2$ to $3.4$ and $\alpha_{\rm ret}$ (Eqn.~\ref{eq:Mreturn}) from $0.64$ to $1.00$. 2) for Gonzalez-Perez et~al., we changed $V_{\rm SN}$ (Eqn.~\ref{eq:Meject}) from $425 \, {\rm km\,  s}^{-1}$ to $380 \, {\rm km \, s}^{-1}$ and $\alpha_{\rm cool}$ (Eqn.~\ref{eq:alpha_cool}) from $0.60$ to $0.72$. Note that, in both cases, we adopt instantaneous ram pressure stripping of hot gas in satellites; the consequences of using instead a gradual ram pressure stripping of the hot halo in satellites are discussed in Appendix~\ref{sec:ram}. 

We note that in all of the PMILL runs we have used all 271 simulation outputs to build the halo merger trees using the {\tt DHALO} algorithm of \cite{Lilian:2014}. The {\tt GALFORM} code can insert  additional timesteps called substeps between the timesteps on which the merger histories are tabulated to improve the accuracy of the calculation of the transfer of mass and metals between various baryon reservoirs and the calculation of the luminosities of galaxies. For the MSI simulation with $\sim 63$ outputs, the model predictions converged with 8 of these additional substeps in time  inserted between the simulation outputs. With four times as many outputs available for the PMILL merger histories we reduced the number of time substeps to 2 to retain the same time resolution in the calculations carried out by {\tt GALFORM}. We note that for the models considered here, the predictions are insensitive to the number of simulation outputs used to construct the halo merger histories (we can use all 271 outputs to construct the halo merger histories, or subsets of these, e.g. 128, 64 etc). The additional N-body snapshots available in the PMILL compared with the WM7 or MSI simulations will, however, reduce the errors introduced by interpolating galaxy positions between snapshots for the construction of catalogues on an observer's past lightcone (see \citealt{Merson:2013}); this aspect will be considered in a separate paper. 

We next consider how the predictions for the HI mass function change when moving from the models run in the WM7 simulation to the PMILL RECAL versions. The top panel of Fig.~\ref{fig:himf} shows that the WM7 model mass functions display a bump at an HI galaxy mass of $M_{\rm HI} \sim 10^{8.5} \, h^{-2} \, {\rm M_{\odot}}$. This feature is partially dependent on the halo mass resolution and shifts by a decade lower in mass to $M_{\rm HI} \sim 10^{7.5} \, h^{-2} \,  {\rm M_{\odot}}$ in the RECAL models (see also \citealt{Power:2010}). The location of this feature also depends upon the photoionisation feedback adopted \citep{Hank:2013,Hank:2015}. All of the models overpredict the abundance of galaxies with $M_{\rm HI} \sim 10^{8} \, h^{-2} \, {\rm M_{\odot}}$ by a factor of approximately three. At the high mass end, the models predict different numbers of galaxies, with the Lacey et~al. RECAL model under predicting the observational estimates and the Gonzalez-Perez et~al. RECAL agreeing with the estimate from  \cite{Zwaan:2005}. We note that the estimates of the HI mass function from the ALFALFA survey by \cite{Martin:2010} and from HIPASS by \cite{Zwaan:2005} are inconsistent with one another within the stated errors at the high mass end. This discrepancy remains following the analysis of the full ALFALFA survey by \cite{Jones:2018}. The model predictions differ from one another by a similar degree to the observational estimates at high masses. 

The more relevant quantity for HI intensity mapping predictions is the mass weighted HI mass function, which illustrates the galaxy masses that make the biggest contribution to the global density of HI. This is quantity is plotted in the lower panel of Fig.~\ref{fig:himf}, which shows that the global density of HI is dominated, as expected, by galaxies around the knee in the mass function, around $M_{\rm HI} \sim 10^{9.7} \, h^{-2} \, {\rm M_{\odot}}$. The model predictions peak within $0.2$ dex of this mass. The low and high mass tails of the mass function contribute relatively little to the HI mass density.

\begin{figure}
	\includegraphics[trim={0.8cm 1cm 0  0.5cm},clip,width=1.05\columnwidth]{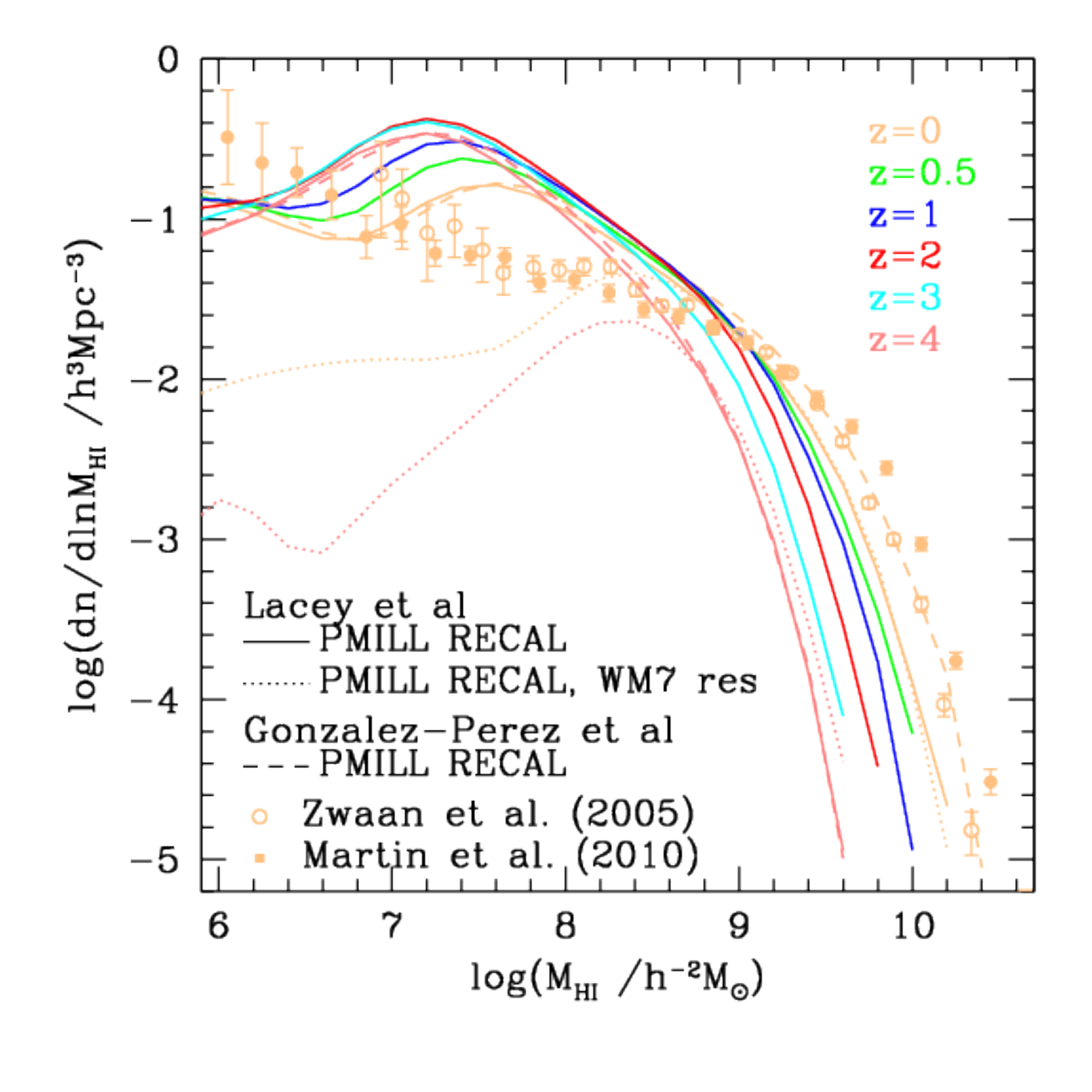}

    \caption{The evolution of the HI mass function. Different colours show different redshifts as labelled. The solid lines show the recalibrated version of the  \citet{Lacey:2016} model. The recalibrated version of the  \citet{GonzalezPerez:2014} is shown by the dashed lines; for clarity these are only plotted at $z=0$ and $z=4$. The dotted lines, which are also only plotted for $z=0$ and $z=4$, show a version of the recalibrated Lacey et~al. model in which the resolution of the halo merger trees is limited to that of the WM7 simulation. The symbols show observational estimates of the mass function at $z=0$ from \citet{Zwaan:2005} and \citet{Martin:2010}.}
    \label{fig:himf_evol}
\end{figure}

\begin{figure}
	\includegraphics[trim={0.5cm 1cm 0  0.5cm},clip,width=1.05\columnwidth]{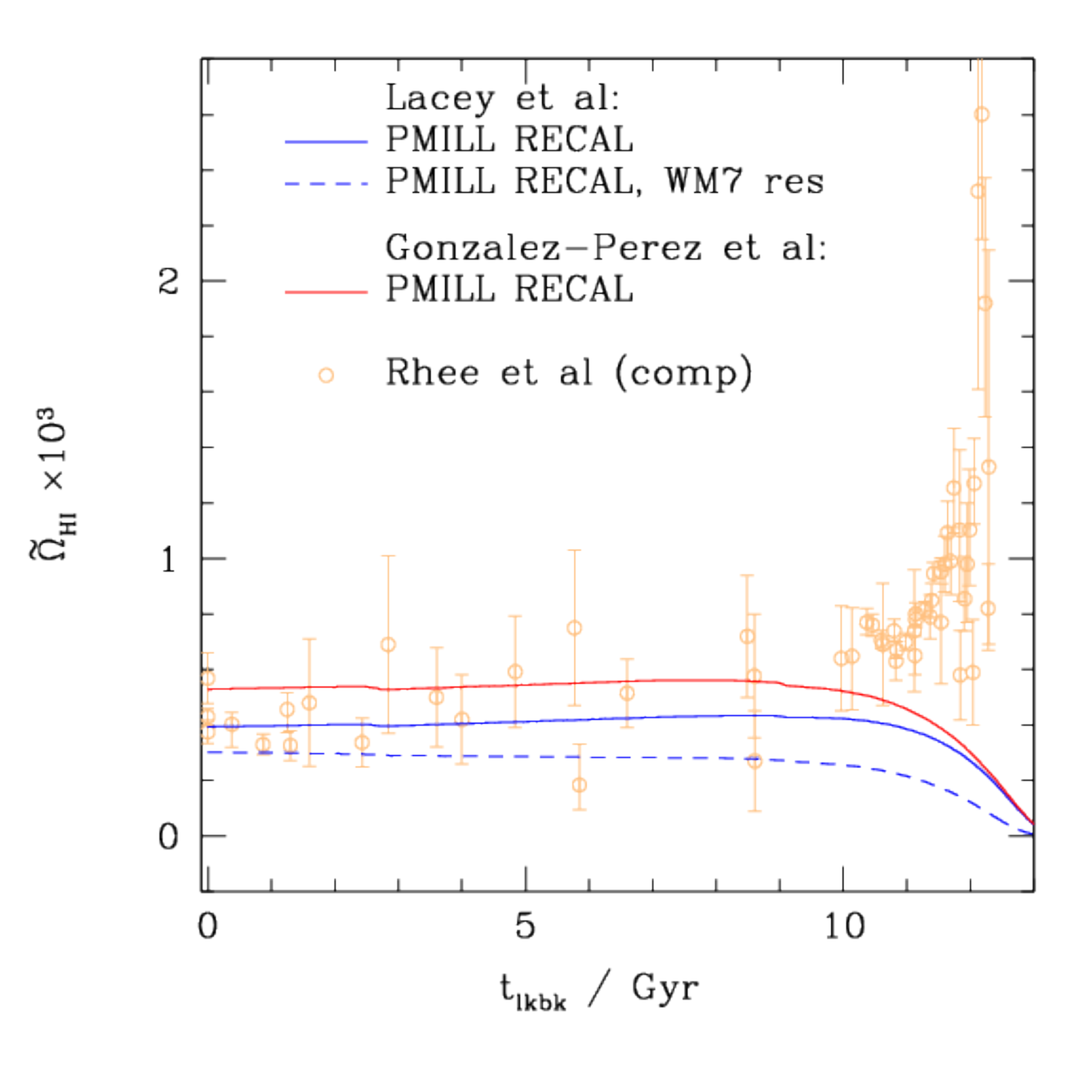}

    \caption{The evolution of the HI density parameter. The solid blue line  shows the recalibrated version of the  \citet{Lacey:2016} model; the dashed blue line shows the results of this model when restricting the halo merger tree resolution to the equivalent of the WM7 simulation. The red solid line shows the prediction of the recalibrated \citet{Gonzalez-Perez:2013} model. The symbols show a compilation of observational estimates taken from Rhee et~al. (2018). Note that the density parameter is in units of the critical density at the present day. }
    \label{fig:omegahi}
\end{figure}

The predicted evolution of the HI mass function in the two recalibrated {\tt GALFORM} models is shown in Fig.~\ref{fig:himf_evol}. For clarity, the Gonzalez-Perez et~al version of the model is only plotted at $z=0$ and $z=4$. The high mass end of the HI mass function declines slowly with increasing redshift to $z=1$, then dropping more rapidly to $z=4$. There is little evolution at intermediate masses until $z>3$. 
The turnover at low masses shifts to lower masses with increasing redshift up to $z=2$. This feature is resolution dependent. The benefit of the improved resolution of the PMILL simulation compared with that of the WM7 run can be seen by rerunning the Lacey et~al version of the recalibrated PMILL model but degrading the resolution of the halo merger trees to the equivalent of the WM7 simulation. The results at $z=0$ and $z=4$ are shown by the dotted lines in Fig.~\ref{fig:himf_evol}. With trees at WM7 resolution, about half of the global density of HI is resolved compared with the predictions at the full PMILL resolution. 

The global density of HI, expressed in units of the critical density today, ${\tilde{\Omega}}_{\rm HI}$, is shown in Fig.~\ref{fig:omegahi}. The two recalibrated models predict similar HI densities at high redshift, and differ only by $\approx 20\%$ at low redshift. The improved halo mass resolution of the P-Millennium simulation is important for resolving more of the global HI density, as can be seen by comparing the  dashed blue curve, obtained from the Lacey et~al. RECAL model run with merger trees limited to the WM7 resolution and the solid blue curve, which shows the results obtained with the full resolution merger trees. 

The models are also compared in Fig.~\ref{fig:omegahi} with a compilation of recent observational determinations of ${\tilde\Omega}_{\rm HI}$ taken from \cite{Rhee:2018} (see references therein for the original sources). The models are in reasonable agreement with these observational estimates over more than 70\% of the history of the Universe but  appear to underpredict the inferred ${\tilde\Omega}_{\rm HI}$ at look-back times in excess of 10 billion years. We note that semi-analytical models have consistently predicted that ${\tilde\Omega}_{\rm HI}$ declines with increasing lookback time  \citep{Lagos:2014,Crighton:2015}. However, an important caveat should be mentioned regarding the comparison between theoretical predictions and the inferences from observations.  The semi-analytical model only considers HI inside galactic disks, and does not account for HI that is within halos but outside the disk, or outside halos altogether \citep{Lagos:2014,Lagos:2018}. The Illustris simulation predicts that a declining fraction of HI is found inside halos with increasing redshift \citep{Villaescusa:2018}. The drop depends on how the halos are defined; in one case as much as a third of the HI could be outside halos by $z=5$. This implies that an increasing fraction of HI in the simulation could be in the outer parts of dark matter halos rather than in a structure resembling a galactic disk. Finally, we note that $\Omega_{\rm HI}$ recovered in post-processing from the Illustris TNG simulations changes relatively little with redshift, though does vary substantial with the resolution of the simulation used \citep{Villaescusa:2018}.

\section{The HI content of dark matter halos}

\begin{figure}
	\includegraphics[trim={0.7cm 1.5cm 0 0},clip,width=1.05\columnwidth]{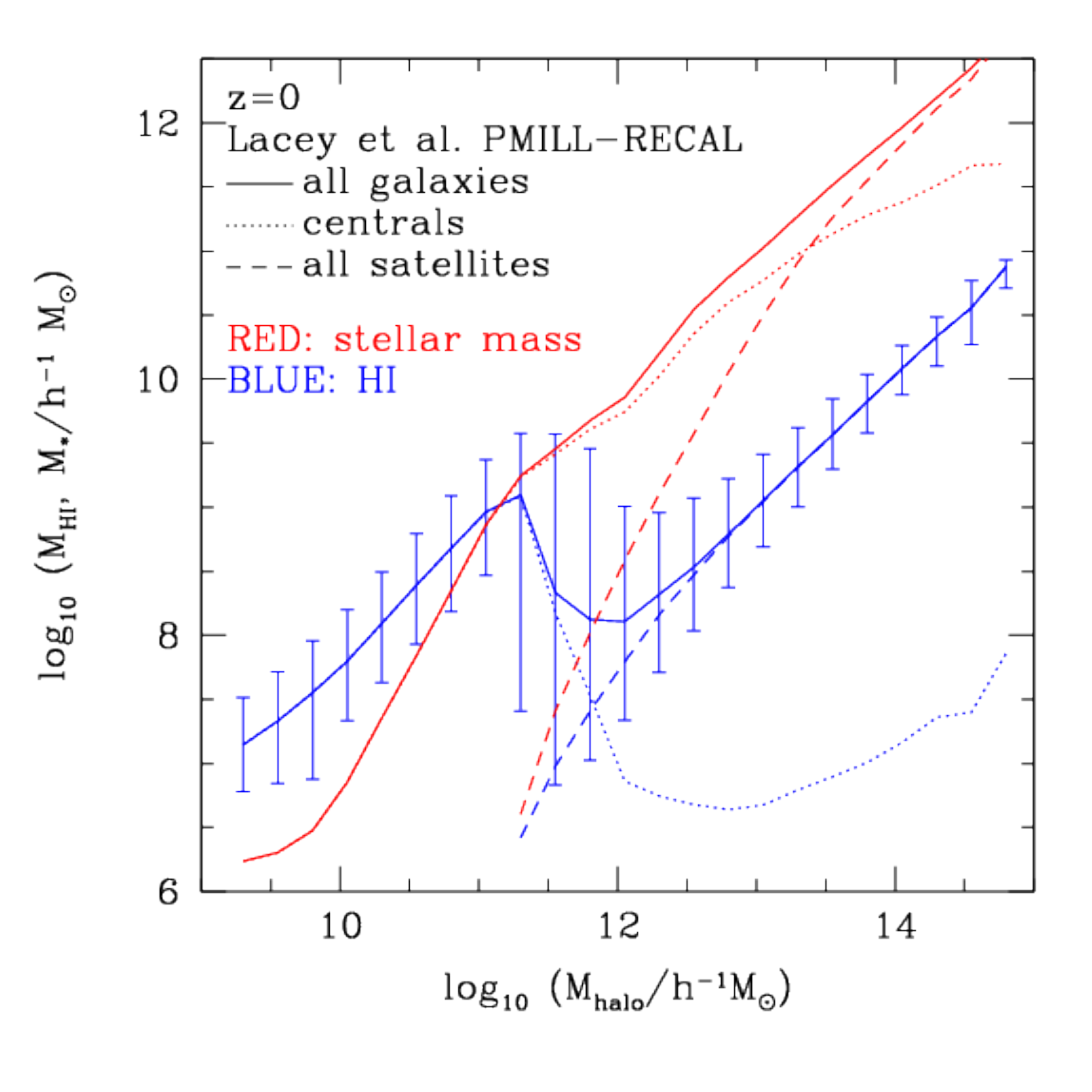}
    	\includegraphics[trim={0.7cm 1.5cm 0 0},clip,width=1.05\columnwidth]{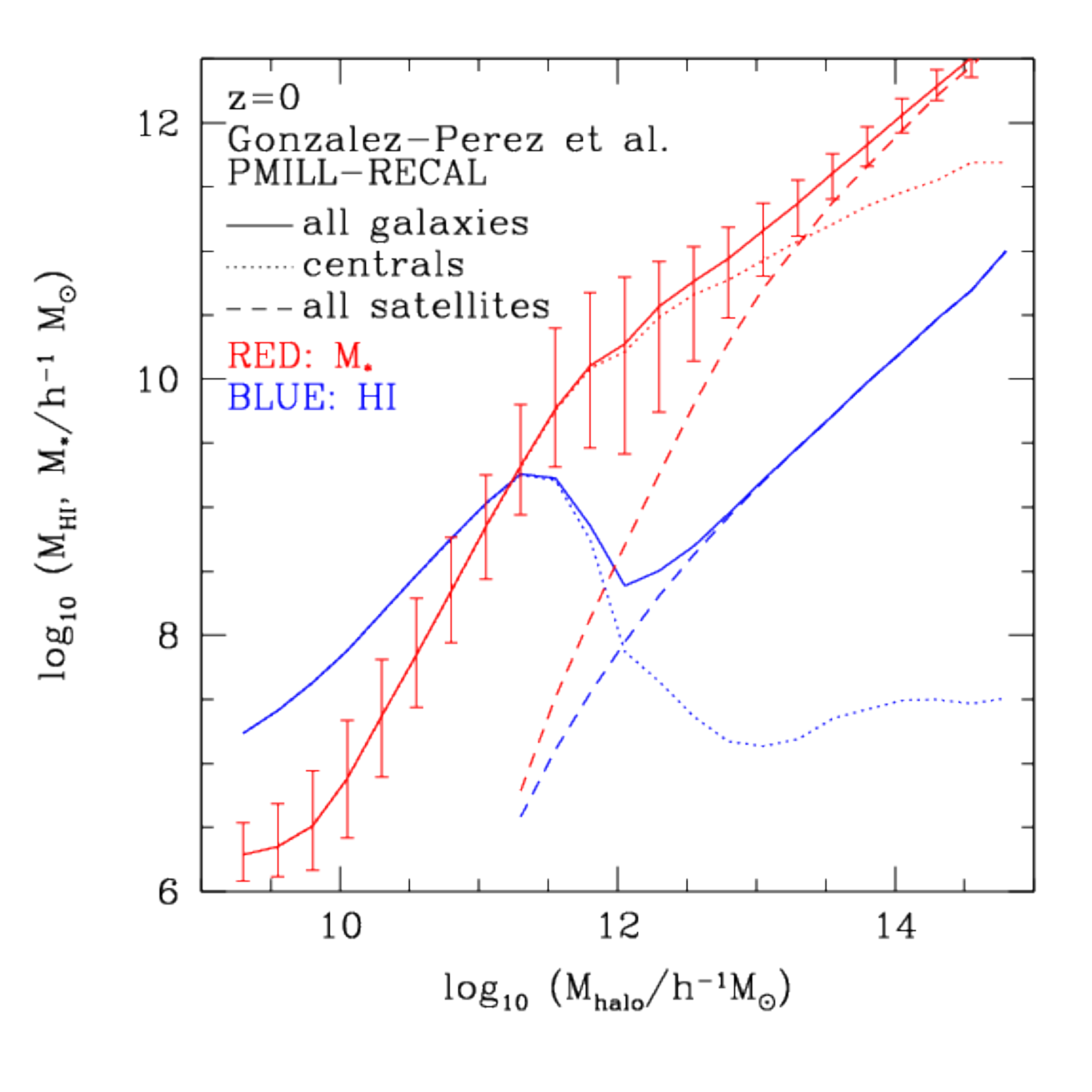}
    \caption{The median HI (blue) and stellar mass (red) contents of dark matter halos in the models as a function of halo mass, showing the contributions from all galaxies (solid), centrals (dotted) and all satellites within a halo (satellites). (Note that the precise value of the median mass for satellite galaxies is resolution dependent.) The bars show the 10-90 percentile range of the distribution, and for clarity are just shown for the total HI  content of the haloes in the top panel and the total stellar mass in the bottom panel.  The top panel shows the predictions of the recalibrated version of the \citet{Lacey:2016} model and the bottom panel shows the recalibrated \citet{GonzalezPerez:2014} model.
    }
    \label{fig:msmhi}
\end{figure}

\begin{figure}
	\includegraphics[trim={0.7cm 1.5cm 0 0},clip,width=1.05\columnwidth]{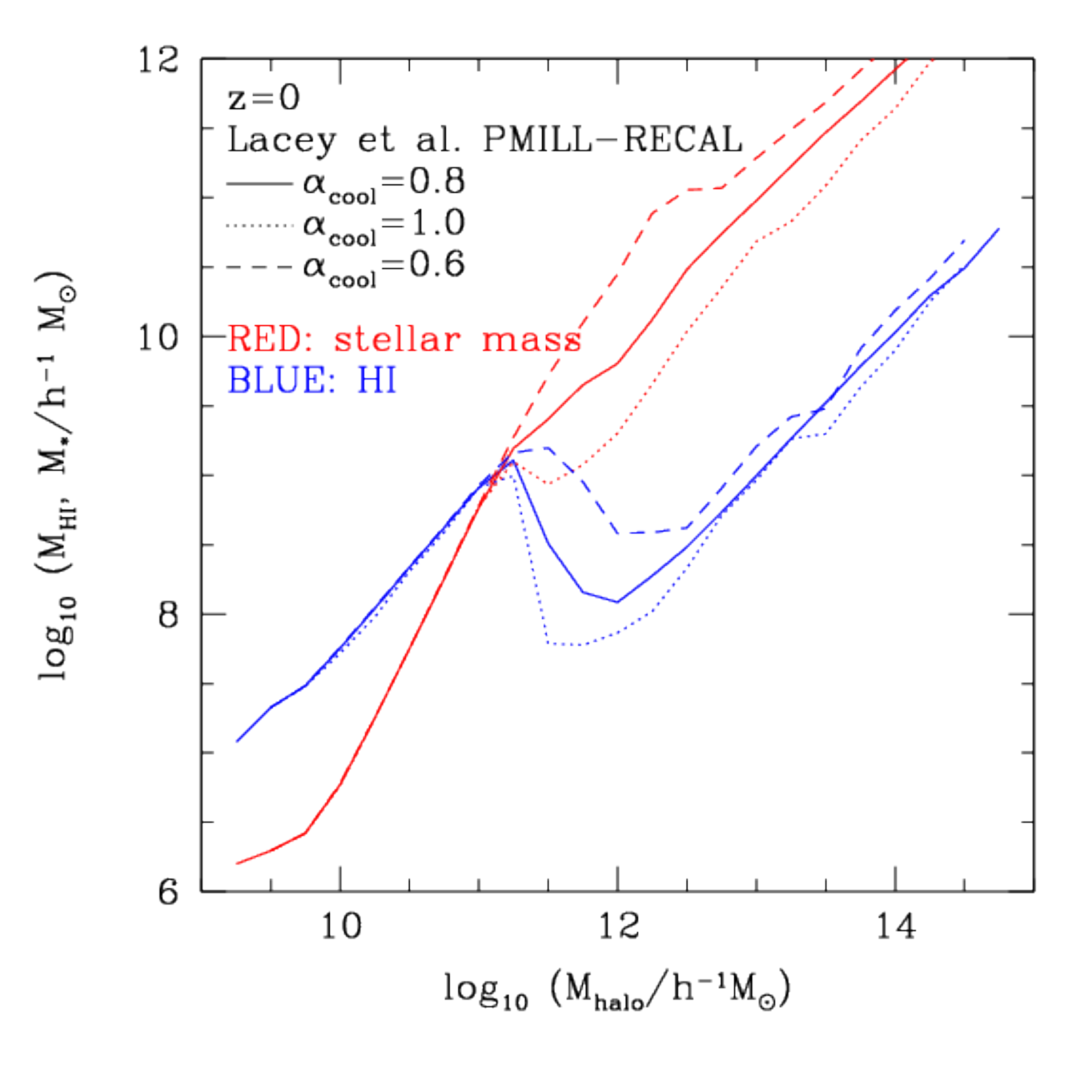}
    	\includegraphics[trim={0.7cm 1.5cm 0 0},clip,width=1.05\columnwidth]{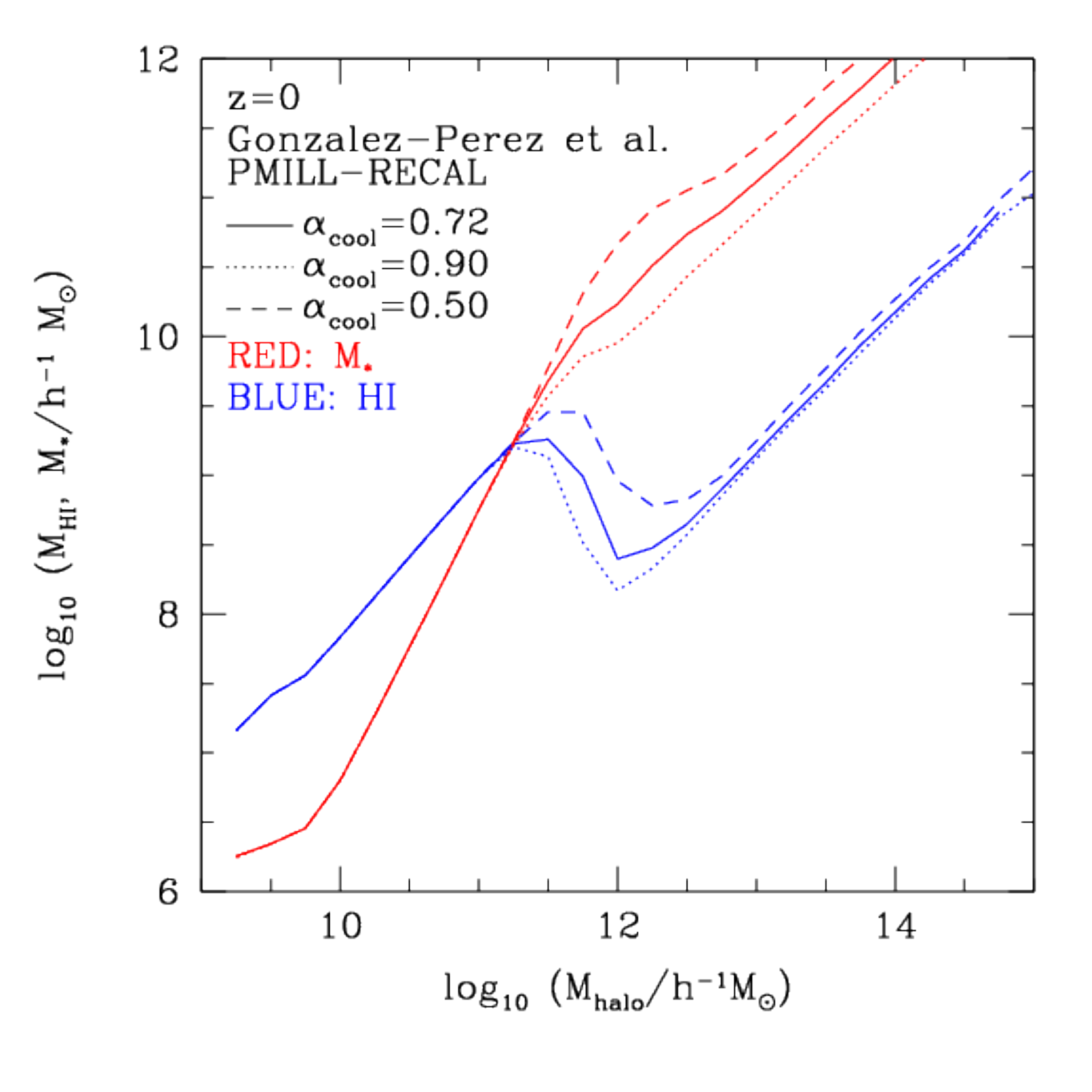}
    \caption{The impact of varying the parameter controlling the onset of AGN feedback, $\alpha_{\rm cool}$ (see Eqn.~\ref{eq:alpha_cool}) on the median total HI (blue) and total stellar mass (red) contents of dark matter halos in the models as a function of halo mass. The different line styles show the predictions for different values of $\alpha_{\rm cool}$, as indicated by the key. The results plotted are for all galaxies in the halo. The predictions for the fiducial values of $\alpha_{\rm cool}$ are shown by the solid lines. The upper panel shows the predictions of the  recalibrated version of the \citet{Lacey:2016} model and the bottom panel shows the recalibrated \citet{GonzalezPerez:2014} model, along with their variants.
    }
    \label{fig:msmhi2}
\end{figure}

With the recalibrated models in hand, we now explore the  predictions for the HI contents of dark matter halos in Fig.\ref{fig:msmhi}. We compare these predictions with the stellar mass contents of halos for reference, even though this quantity does not readily lend itself to an observational test. A plot similar to Fig.~\ref{fig:msmhi} was made by \cite{Hank:2017} for a modified version of the \cite{Lagos:2012} model run in the MSII simulation. 

Fig.\ref{fig:msmhi} shows that the median HI mass of central galaxies tracks the mass of their  host dark matter halo between $M_{\rm halo} \sim 10^{9} \, h^{-1} \,  {\rm M_{\odot}}$ and $10^{11} \, h^{-1} \,  {\rm M_{\odot}}$, scaling as $M_{\rm HI} \propto M_{\rm halo}^{1.5}$. There is a strong break above $10^{11} \, h^{-1} \, M_{\odot}$ due to AGN feedback (as we demonstrate below). This break occurs at slightly different masses in the two models due to the different values adopted for the AGN feedback parameter,  $\alpha_{\rm cool}$ (see Eqn.~\ref{eq:alpha_cool}). Above this halo mass there is a dramatic drop in the HI mass of central galaxies. This is due to a similar drop in the cold gas mass due to the suppression of gas cooling in these halos (see \citealt{Hank:2011}). Cold gas is brought into central galaxies in these halos only by galaxy mergers. Fig.~\ref{fig:msmhi} shows that the median HI mass of the total satellite population in each halo is remarkably similar in the RECAL versions of the Gonzalez-Perez et~al. and Lacey et~al. models. \footnote{Note that both these recalibrated models assume that galaxies are fully stripped of their hot gas halos when they become satellites; we explore an alternative model with gradual ram pressure stripping of the hot gas in Appendix~\ref{sec:ram}.} There is a substantial scatter, particularly around the break. The predictions from {\tt GALFORM} tend to show more scatter in galaxy properties at a given halo mass than other semi-analytical models such as, for example, {\tt L-GALAXIES} or gas dynamical simulations such as {\tt EAGLE} \citep{Contreras:2015,Guo:2016,Zoldan:2017}. One explanation for this is the treatment of supernova feedback \citep{Guo:2016,Mitchell:2016}. In {\tt GALFORM} the bulge and disk components of a galaxy could experience different mass loading of the winds driven by supernovae, due to the different circular velocities calculated for the disk and bulge that are used in Eqn~\ref{eq:Meject} (see \citealt{Cole:2000} for a description of the calculation of the circular velocity of the disk and bulge, which assumes conservation of the angular momentum of the cooling gas and takes into account the gravity of the baryons and their effect on the dark matter halo). In {\tt L-GALAXIES}, the disk and bulge experience the same supernova wind as the halo circular velocity is used in the prescription used to set the mass-loading of the SNe driven wind.

The HI content of dark matter halos is dominated by central galaxies until a halo mass of $M_{\rm halo} \sim 10^{11} - 10^{11.5} \, h^{-1} \, {\rm M_{\odot}}$, after which there is a sharp dip in the HI content until the satellites dominate the HI content for $M_{\rm halo} > 10^{12} \, h^{-1} \, {\rm M_{\odot}}$. The slope of the HI mass -- halo mass relation is remarkably similar at low and high halo masses, either side of the kink which marks the onset of AGN feedback. 

In contrast to the behaviour of the HI mass with halo mass, Fig.~\ref{fig:msmhi} shows that the stellar mass of centrals has a stronger dependence on host halo mass, $M_{*} \propto M^{2.2}_{\rm halo}$ (see \citealt{Mitchell:2016}), up until the halo mass where AGN feedback becomes important. The change in slope of the stellar mass - halo mass relation arguably happens at a slightly higher halo mass than it does for the HI mass - halo mass relation, particularly in the recalibrated Gonzalez-Perez et~al. model. The stellar mass  - halo mass relation in the Lacey et~al. model shows a feature around this mass, which makes it difficult to locate the change in slope. Also, the halo mass at which the satellite population dominates over the central galaxy is much higher for the case of stellar mass ($M_{\rm halo} \approx 2 \times 10^{13} h^{-1} {\rm M_{\odot}}$) than it is for HI mass ($M_{\rm halo} \approx 10^{12} h^{-1} {\rm M_{\odot}}$).

The position and form of the break in the scaling of the stellar mass and HI mass contents of halos with halo mass is the result of the interplay between a number of processes: gas cooling, gas heating by supernovae and AGN, star formation, galaxy mergers and the timescale for gas heated by supernovae to be reincorporated into the hot gas halo (see \citealt{Lacey:2016} for illustrations of how the model predictions depend on varying the parameters that govern these effects). One of the attractive features of semi-analytical modelling is that we can vary the value of a parameter to see the effect this has on the model predictions. We caution the reader that these variant models are purely illustrative and should not be viewed as viable models, since they do not satisfy the observational tests required of fiducial models. Fig.~\ref{fig:msmhi2} shows how the model predictions for the stellar mass and HI contents of halos respond to perturbing the value of $\alpha_{\rm cool}$. The change in slope of the stellar mass $-$ halo mass relation moves to higher halo mass on reducing the value of $\alpha_{\rm cool}$ (reducing the value of the parameter $\alpha_{\rm cool}$ shifts the onset of AGN feedback to more massive haloes); the reduction in the break mass on increasing $\alpha_{\rm cool}$ is less substantial. The changes in the halo mass at which the turn-over in the HI mass $-$ halo mass relation occurs are less dramatic then those in the case of stellar mass, but are nevertheless in the same sense. 

\begin{figure}
	\includegraphics[trim={0cm 3 0 0},clip,width=1.05\columnwidth]{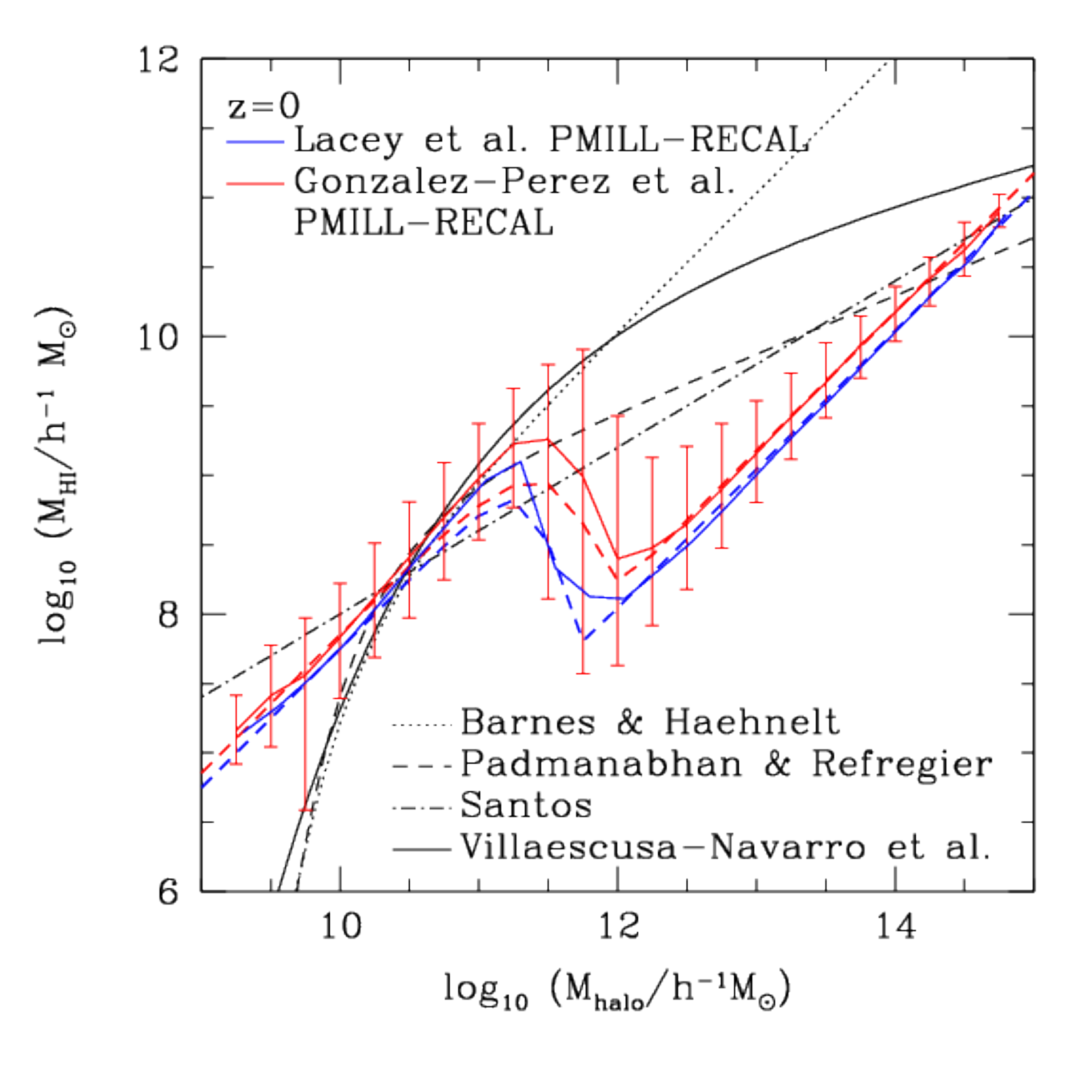}
    \caption{The median total HI content as a function of halo mass at $z=0$. The solid lines show the predictions of the recalibrated versions of the \citet{Lacey:2016} (blue) and \citet{GonzalezPerez:2014} (red) models in the P-Millennium run. The dashed lines of the same colour show the best fitting version of the parametric model describing these results, using Eqn.~\ref{eq:fit} in the text. The black lines show selected empirical fits from the literature: \citealt{Barnes:2010} (dotted), \citealt{Padmanabhan:2017a} (dashed),  \citealt{Santos:2015} (dot-dashed) and Villaescusa-Navarro et~al. (2018) (solid line); the latter is a fit to the Illustris TNG-100 simulation}.
    
    \label{fig:mhimhalofit}
\end{figure}

\begin{figure}
	\includegraphics[trim={0cm 2 0 0},clip,width=1.05\columnwidth]{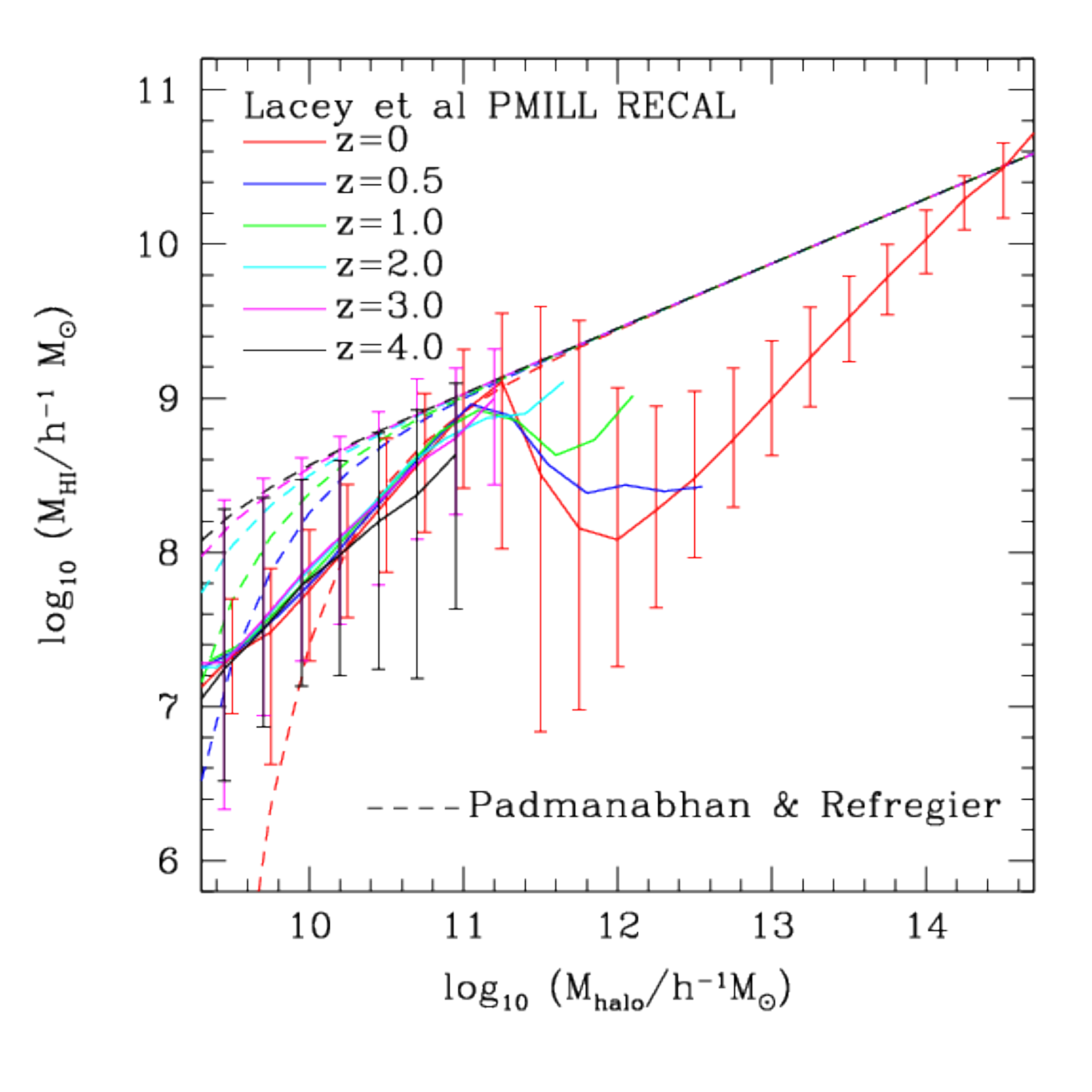}
    \includegraphics[width=1.05\columnwidth]{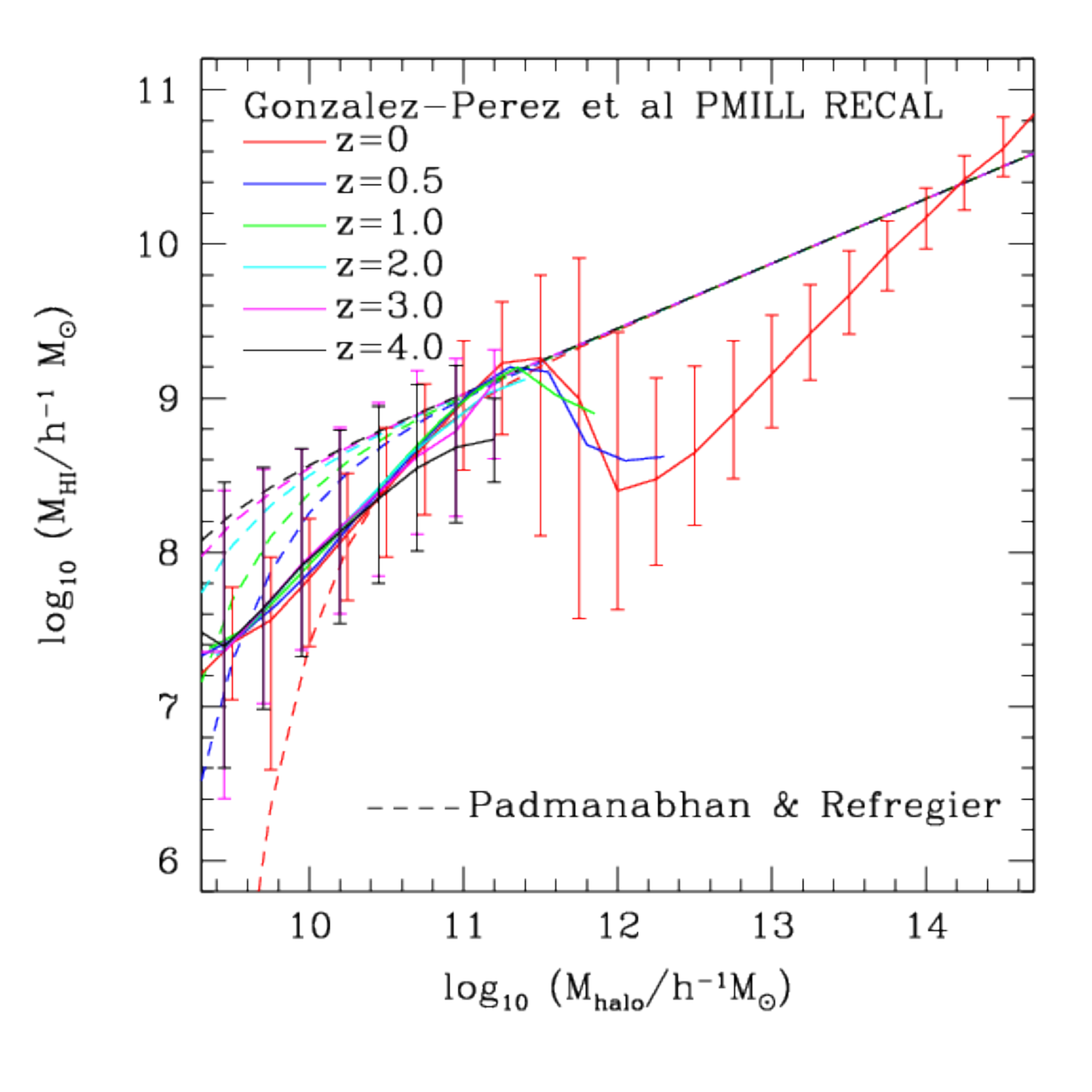}
    \caption{
    The median total HI content as a function of halo mass, showing the contribution from all galaxies. The top panel shows the predictions for the Lacey et~al. PMILL RECAL model and the bottom panel shows the Gonzalez-Perez et~al. RECAL model. The different colours show the predictions at different redshifts, as indicated by the legend. The bars show the 10-90 percentile range of the distribution and are plotted for the predictions at $z=0$ and $z=3$.The dashed lines show the empirical HI mass -- halo mass model from Padmanabhan \& Refregier (2017a) at different redshifts, following the colour key.
    }
    \label{fig:mhimhaloz}
\end{figure}

Many authors have proposed empirical models to describe the HI mass $-$ halo mass relation  \citep{Barnes:2014,Santos:2015,Padmanabhan:2015,Popping:2015Padmanabhan:2016,Padmanabhan:2017a,Padmanabhan:2017b,Villaescusa:2018,Obuljen:2018}. We compare a small selection of these against the predictions of the semi-analytical models in Fig.~\ref{fig:mhimhalofit}.  The \cite{Santos:2015} model is a simple power law in halo mass that is generally shallower than the slope of the semi-analytical model predictions. None of the empirical models are designed to allow for a break feature around the halo mass where AGN feedback first suppresses gas cooling in the semi-analytical models. The form of the HI-mass -- halo mass relation from \cite{Padmanabhan:2017a} displays more curvature than that predicted by the semi-analytical models at low halo masses, particularly at $z=0$. \cite{Barnes:2014} also argued for a steep HI mass -- halo mass relation at low halo masses at $z=0$, based on fits to the observed abundance of damped Lyman-$\alpha$ absorbers (see also \citealt{Barnes:2009,Barnes:2010}). \cite{Villaescusa:2018} gave a fit to the HI content of dark matter halos identified using a friends-of-friends algorithm in the Illustris TNG100 simulation, after post-processing the cold gas mass to estimate the HI mass. Similar to the empirical studies discussed above, the Illustris results also do not contain a break feature. We discuss the implications of this for our results in the Conclusions. A similar fit to that used by Villaescusa-Navarro et~al. was used to describe the HI content of halos in the ALFALFA survey by \cite{Obuljen:2018}.

Our model predictions for the total HI mass in halos can be parametrized as 

\begin{equation}
\frac{M_{\rm HI}}{M_{\rm halo}} = 
A_{1} 
\exp \left[ -\left( \frac{M_{\rm halo}}{M_{\rm break}}\right)^\alpha \right] \times  
\left( \frac{M_{\rm halo}}{10^{10} \, h^{-1} \, {\rm M_{\odot}}} \right)^\beta + A_{2},  
\label{eq:fit}
\end{equation}
where $(A_{1}, \, A_{2}, \, M_{\rm break}, \, \alpha, \beta)$ are parameters. This parametrisation assumes that $M_{\rm HI} \propto M_{\rm halo}^{1 + \beta}$ at low halo masses and $\propto M_{\rm halo}$ at high halo masses. The parameter $\alpha$ controls the sharpness of the break in the relation at halo mass $M_{\rm break}$. For the Lacey et~al. PMILL RECAL, $(A_{1}, A_{2}, M_{\rm break}, \alpha, \beta) = (0.0055, 1.1 \times 10^{-4}, 10^{11.4} h^{-1} {\rm M_{\odot}}, 2.5, 0.2)$ and for the Gonzalez-Perez et~al. RECAL $(A_{1}, A_{2}, M_{\rm break}, \alpha, \beta) = (0.007, 1.5 \times 10^{-4}, 10^{11.5} h^{-1} {\rm M_{\odot}}, 1.5, 0.2)$. These fits are shown by the blue and red dashed lines in Fig.~\ref{fig:mhimhalofit}. We caution the reader that this relation should only be used down to the halo mass resolution limit of the P-Millennium, $2.12 \times 10^{9} h^{-1} \, {\rm M_{\odot}}$.

The predicted evolution of the HI content of dark matter haloes is plotted in Fig.~\ref{fig:mhimhaloz}. There is remarkably little change in the relation predicted by the models between $z=0$ and $z=3$. The main effects are a depopulation of the high halo mass part of the relation due to the hierarchical growth of the halo mass function and some minor variation in the form of the relation around the feature that arises due to the onset of AGN feedback.  

The lack of evolution in the semi-analytical model predictions for the HI mass -- halo mass relation is in stark contrast to that displayed by empirical models taken from the literature, examples of which are also plotted Fig.~\ref{fig:mhimhaloz}. The model proposed by \cite{Padmanabhan:2017a}, which is constrained to reproduce various measurements of the HI content of the observed galaxy population over a range of redshifts, displays a substantial increase in the HI-content of low mass haloes between $z=0$ and $z=0.5$, with more modest evolution thereafter to $z=3$. The HI mass $-$ halo mass relation does not evolve at high halo masses, for which the available observational data do not constrain the model parameters.

\section{Clustering}

\begin{figure}
	\includegraphics[trim={0.2cm 4 0 1},clip,width=1.08\columnwidth]{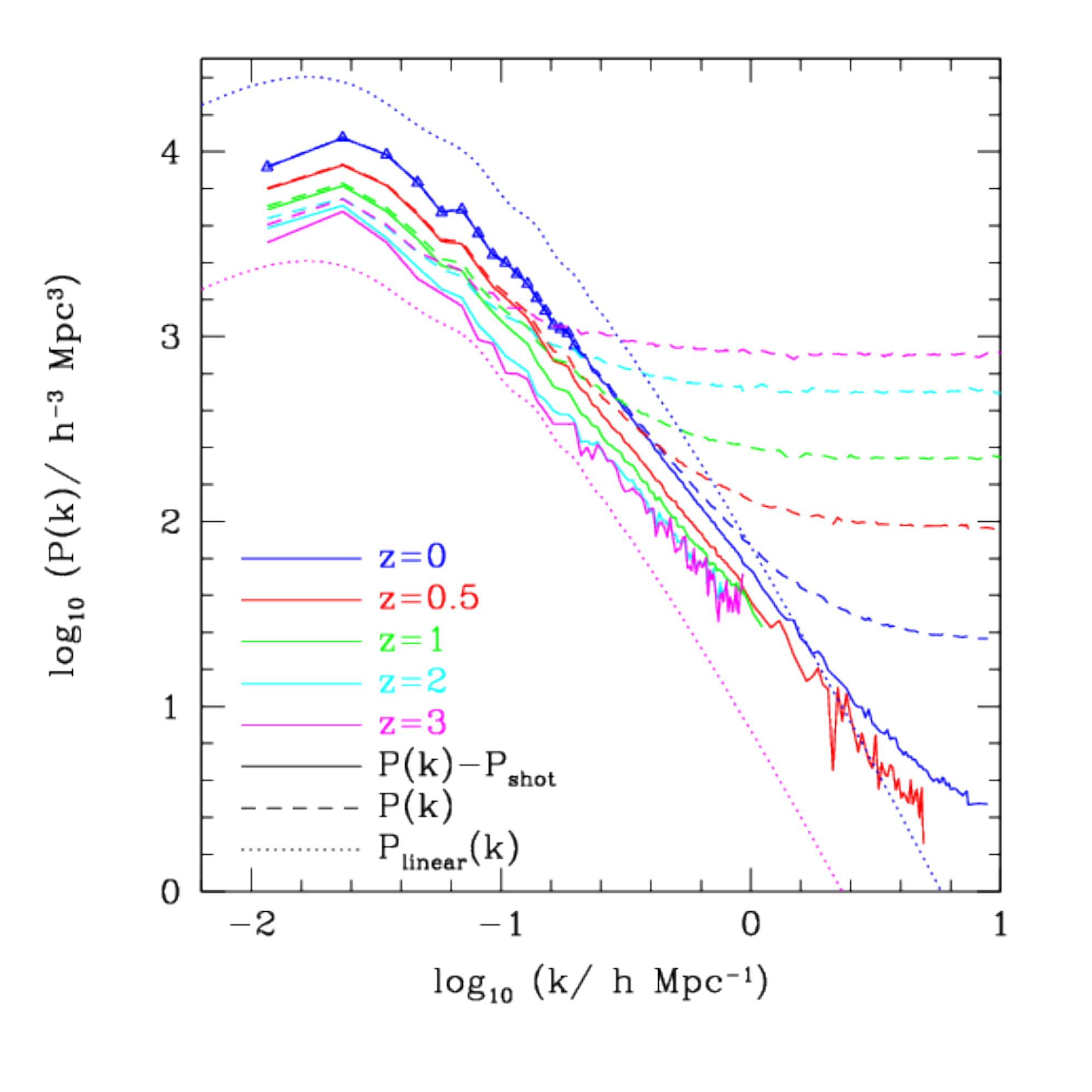}
    \caption{The power spectrum of HI sources as a function of redshift, with different line colours and line styles indicating the redshift as labelled. The dashed lines show the measured power spectrum and the solid lines show the power spectrum after subtracting Poisson shot noise. The dotted lines shows the linear perturbation theory power spectrum plotted at $z=0$ (blue) and $z=3$ (magenta) to serve as a reference. The power spectrum corrected for shot noise is only plotted at wavenumbers for  which the spectrum is not excessively noisy. The bin positions at low wavenumbers are indicated by symbols on the $z=0$ curve.
    }
    \label{fig:pk}
\end{figure}

\begin{figure}
	\includegraphics[trim={0.2cm 4 0 1},clip,width=1.08\columnwidth]{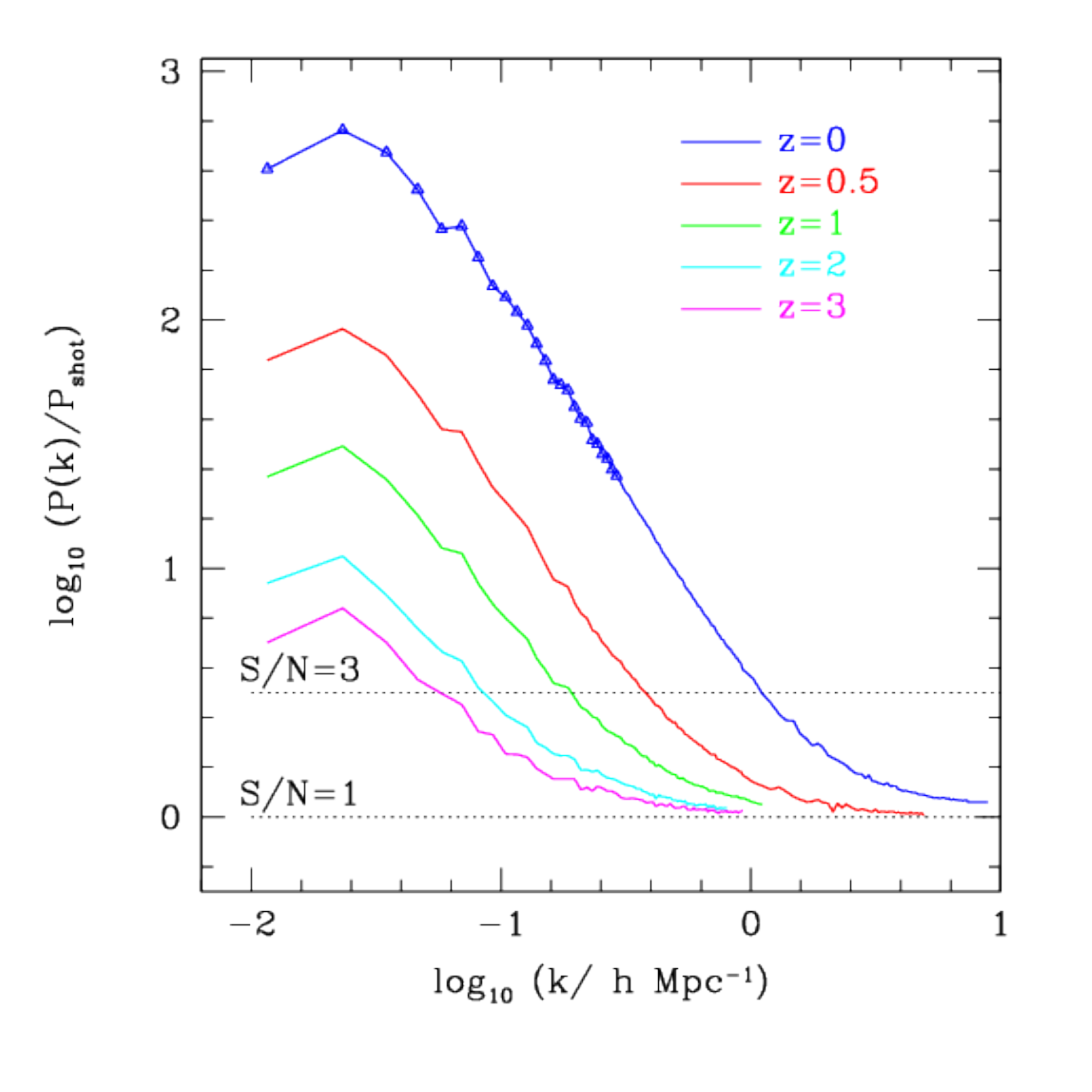}
    \caption{The signal-to-noise ratio of the power spectrum of HI sources, defined as the measured power spectrum in units of the shot noise, with different line colours and line styles indicating the redshift as labelled. A signal-to-noise ratio 3 is often considered as a lower limit for a statistically useful measurement of the power spectrum. 
    }
    \label{fig:sn}
\end{figure}

\begin{figure}
	\includegraphics[trim={0.2cm 4 0 1},clip,width=1.08\columnwidth]{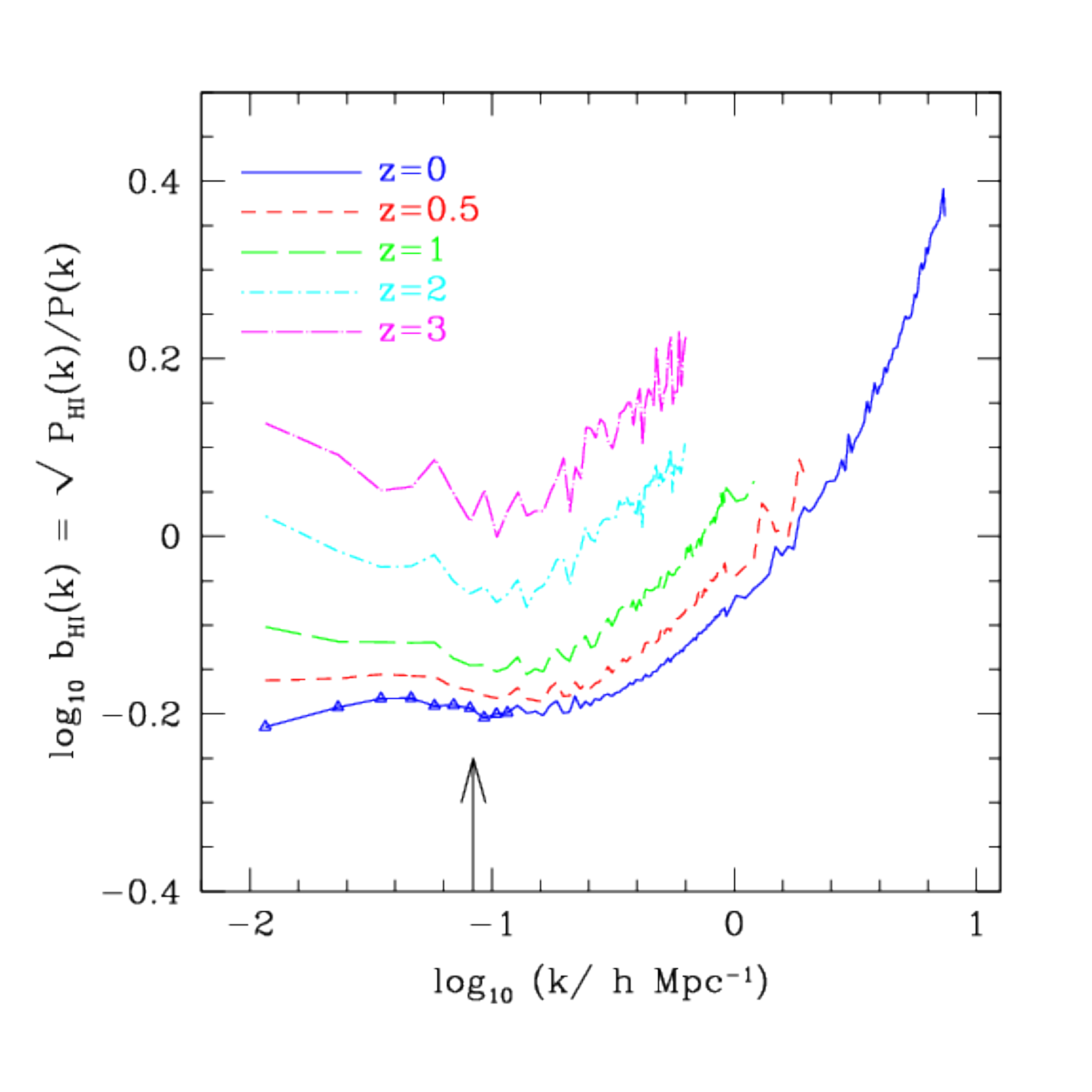}
    \caption{The bias of HI sources as a function of redshift, with different line colours and line styles indicating the redshift as labelled. The power spectrum of HI sources is estimated in real space after placing the entire HI content of each halo at its centre of mass. The shot noise is subtracted from the power spectrum before computing the bias. The linear power spectrum is estimated directly from the simulation on large scales to remove the effects of sampling variance. The binning of the power spectrum in this case in indicated by the symbols plotted on the $z=0$ curve. At higher wavenumbers the analytic linear theory power spectrum is used. The wavenumber corresponding to the box size of the Illustris TNG100 simulation is indicated by the arrow, which is the largest scale probed by the measurements of \citet{Villaescusa:2018}.
    }
    \label{fig:bias}
\end{figure}

By implementing {\tt GALFORM} in the P-Millennium N-body simulation we are able to make a direct prediction of the clustering of HI sources and their bias compared to the underlying dark matter distribution. 

We estimate the power spectrum of HI sources that is relevant for intensity mapping predictions by assigning a weight to each halo that is equal to the total mass of HI contained in galaxies hosted by the halo. This is the weight that is assigned to the density grid used to estimate the power spectrum, using the centre of mass position of the halo in real space i.e. without taking to account the impact of peculiar velocities. The power spectrum measured at different redshifts is plotted in Fig.~\ref{fig:pk}. The dashed lines show the measured power spectrum without any corrections for shot noise. The shot noise is estimated in the simulation as 
\begin{equation}
P_{\rm shot} = L^{3}_{\rm box} \frac{\sum_{\rm i} w^{2}_{\rm i}}{\sum_{\rm i} w_{\rm i}}, 
\end{equation}
where $L_{\rm box}$ is the length of the side of the simulation cube, $w_{i}$ is a weight for each halo that is equal to the total HI mass in the halo and the summation is over all dark matter halos. 
Observationally, the shot noise could be inferred from the asymptotic value of the measured power spectrum at high wavenumbers, as the uncorrected power spectrum tends to the shot noise level on these scales as shown by Fig.~\ref{fig:pk}.
\cite{Villaescusa:2018} showed that the shot noise is significant on small scales when placing the total HI content of the halo at the centre of the halo in this way. Our results show that the shot noise increases with redshift, whereas \citet{Villaescusa:2018} found that the shot noise measured from the Illustris TNG100 simulation increased to $z=1$ before declining strongly to higher redshifts. We note that this difference could be due in part to the difference in mass resolution between the Illustris TNG100 simulation and the P-Millennium. If we restrict our catalogue to halos with 100 particles or more rather than a minimum of 20 particles, a factor of five change in halo mass, then the shot noise we measure at $z=0$ increases by only 20\%. The amplitude of the measured power spectrum at smaller wavenumbers, corresponding to larger length scales, declines little more than a factor of two over the redshift range considered. This is a much weaker change than that in the matter power spectrum, which changes in amplitude by a factor of nine between $z=0$ and $z=3$.  After removing the shot noise, the power spectrum shape does not change much with redshift. 

The visibility or signal-to-noise expected in the power spectrum measurements is illustrated in Fig.~\ref{fig:sn}, which shows the predicted power spectrum in units of the shot noise at different redshifts. Note that in this idealised calculation, we are not considering the smoothing in angular scale or frequency that would be made in a HI intensity mapping measurement, so the Poisson shot noise is the dominant source of noise in our measurement of the power spectrum. In forecasts of cosmological constraints from power spectra, the power spectrum plotted in these units is typically assumed to have a value of at least $\approx 3$ \cite{DETF:2006}. Fig.~\ref{fig:sn} shows that the range of wavenumbers over which this conditions holds is reduced with increasing redshift, due to the slight drop in the amplitude of the power spectrum and the increase in the shot noise. 

The implications of the trends in the power spectrum described above for the bias of HI sources are shown in Fig.~\ref{fig:bias}. This plot shows the bias obtained by taking the measured power spectrum, with shot noise subtracted, and dividing by the corresponding linear perturbation theory redshift. The bias of halos weighted by their HI content is approximately constant on large scales (small wavenumbers) at low redshift. The scale dependence of the bias becomes apparent at smaller wavenumbers with increasing redshift. Similar conclusions have been reached in analyses of the Illustris simulations by \citet{Villaescusa:2018} and \citet{Ando:2018}; the maximum scale measurable in those simulations is shown by the arrow in Fig.~\ref{fig:bias}. Recently, \cite{Obuljen:2018} reported a bias of $0.875$ from a measurement of the clustering of ALFALFA sources; our prediction for the bias is somewhat lower but corresponds to lower mass galaxies.

\section{Summary and Conclusions}

We have presented implementations of the {\tt GALFORM} models introduced by \cite{GonzalezPerez:2014} and \cite{Lacey:2016} in a new, high resolution N-body simulation, the P-Millennium or PMILL run. The models required a minor recalibration due to the improved mass resolution of the halo merger trees extracted from the PMILL, compared to those available from the simulation originally used to calibrate the models. The change in cosmology from a WMAP-7 model to a Planck cosmology does not require a significant change in the model parameters. We also took this opportunity to update the treatment of galaxy mergers in {\tt GALFORM}, using the model introduced by \cite{Simha:2017}. In the end, only minor changes were required to two model parameters in each case to obtain a similar level of agreement with the observational data used to set the model parameters.   

One clear application of the improved halo mass resolution in the P-Millennium is to make predictions for the atomic hydrogen content of dark matter haloes. Observational determinations of the HI mass function are in their infancy and show significant disagreement at high masses. Nevertheless, current estimates do agree with one another around the break in the mass function and suggest that the global HI density at the present day is dominated by galaxies with HI masses $\sim 10^{9.5} h^{-2} {\rm M_{\odot}}$. Our model predictions show that these are central galaxies in halos with mass $\approx 10^{11.5} h^{-1} {\rm M_{\odot}}$, which is approximately the halo mass above which the suppression of gas cooling by AGN heating becomes important in the models. In the PMILL, such haloes are resolved by $\sim 3000$ particles and have reliable merger histories. A calculation made using the same galaxy formation parameters but with dark matter halo merger trees restricted to the same mass resolution as the WM7 simulation resolves around half of the global HI mass recovered at the PMILL resolution. 

There have been a large number of recent studies that have proposed empirical parametric forms for the HI mass $-$ halo mass relation, which is a key input for HI intensity mapping predictions (\citealt{Santos:2015, Padmanabhan:2015,Padmanabhan:2016,Padmanabhan:2017a,Padmanabhan:2017b}). Our ab initio predictions show clear differences from the results of those studies. We find a sharp break in the HI mass $-$ halo mass relation above the halo mass for which AGN heating stops gas cooling onto central galaxies. Also, the form of the predicted relation shows remarkably little dependence on redshift over the interval $z=0-3$. The break in the HI mass $-$ halo mass relation marks a shift from central galaxies dominating the HI content of low mass halos, to the combined satellite population becoming more important in high mass halos. The depth of the dip at the break is reduced somewhat if the gradual ram pressure stripping of the hot gas halos of satellite galaxies is allowed. 

It is not straightforward to compare the predictions of our model to others in the literature, as most calculations follow the total cold gas mass rather than considering the atomic and molecular hydrogen contents of galaxies separately. For example, \cite{Martindale:2017} calculated the HI mass of galaxies in the {\tt L-GALAXIES} model in post-processing and used the HI mass function as a constraint on the model parameters. This resulted in an improved fit to the low mass end of the HI mass function, compared to the prediction of the \cite{Henriques:2015} model. As demonstrated by \cite{Lagos:2011b}, however, post-processing model predictions to compute the HI mass of galaxies can give very different predictions to self-consistently changing the star formation law and computing the evolution of the atomic and molecular hydrogen contents of galaxies. A small number of models do track the atomic and molecular hydrogen contents of galaxies self consistently (\citealt{Fu:2010, Lagos:2011, Lagos:2012, Popping:2014, Xie:2017,Stevens:2017,Lagos:2018}).

The strength of the break in the HI mass $-$ halo mass relation could be sensitive to the way in which different processes are modelled in {\tt GALFORM}, such as AGN feedback or the cooling of gas in satellite galaxies. The treatment of gas cooling in satellites and its impact on the model predictions is discussed in Appendix~\ref{sec:ram}. Regarding the modelling of AGN feedback in {\tt  GALFORM}, once a halo satisfies the conditions for AGN heating to affect gas cooling (see Eqns.~\ref{eq:alpha_cool} and \ref{eq:f_Edd}), the cooling flow is turned off completely. In the {\tt L-GALAXIES} semi-analytical model, for example, the suppression of cooling sets in more gradually  (\citealt{Croton:2006,Henriques:2017}). \cite{Zoldan:2017} compared the HI mass $-$ halo mass relations predicted by a range of semi-analytical models, including an earlier version of the {\tt GALFORM} model by \cite{Bower:2006}. Zoldan et~al. made a similar plot to our Fig.~\ref{fig:himhalo1}, but did not go on to examine  the total gas content of dark matter haloes. The comparison of Zoldan et~al. shows that, out of the models considered, AGN feedback is most efficient at stopping gas cooling in the Bower et~al. model. Nevertheless, the typical  satellite masses are very similar between models, suggesting a total HI mass $-$ halo mass relation that would be similar to the one presented here. 

Recently, hydrodynamic simulations of cosmologically representative volumes have been able to reproduce the observed stellar mass function and other observables  \citep{Vogelsberger:2014, Schaye:2015}. \cite{Crain:2017} present predictions for the HI content of galaxies in the EAGLE simulation of \cite{Schaye:2015}. The HI masses are calculated in post-processing. Crain et~al. state that the predictions for the HI mass function are poorly converged, with a substantial change on improving the mass resolution. Curiously, the fiducial EAGLE run does reproduce the cosmic abundance of HI with redshift reasonably well, despite not matching the present day HI mass function \citep{Rahmati:2015}. This is an observable that semi-analytical models tend to struggle to match at $z>0$ \citep{Popping:2014, Crighton:2015}. \cite{Villaescusa:2018} present predictions for the HI content of halos by post-processing the Illustris TNG simulation described by \cite{Nelson:2018}. Villaescusa et~al. find that there is no break in the HI $-$ halo mass relation, although the slope does get shallower for halos in which AGN feedback is important. These authors also find that satellites dominate the HI content of massive halos. However, the mass at which this transition occurs is somewhat higher than in our predictions. 

The predictions presented here extend those of \cite{Hank:2017}, who considered a version of the \cite{Lagos:2012} model with a new treatment of the suppression of gas cooling in low mass haloes due to photoionisation heating of the intergalactic medium. The model presented here uses a recently determined cosmology and a higher resolution N-body simulation, and has been recalibrated to reproduce selected observations of the galaxy population. The predictions that we have presented for the HI mass $-$ halo mass relation and its evolution will help to guide forecasts for the performance of HI intensity mapping experiments to probe the nature of dark energy.

\section*{Acknowledgements}
We acknowledge conversations with Chris Power, Hank Kim and Qian Zheng. CMB acknowledges research leave granted by Durham University. This work was supported by the Science and Technology Facilities Council [ST/P000541/1].  This work used the DiRAC Data Centric system at Durham University, operated by the Institute for Computational Cosmology on behalf of the STFC DiRAC HPC Facility (www.dirac.ac.uk). This equipment was funded by BIS National E-infrastructure capital grant ST/K00042X/1, STFC capital grant ST/H008519/1, and STFC DiRAC Operations grant ST/K003267/1 and Durham University. DiRAC is part of the National E-Infrastructure.




\bibliographystyle{mnras}
\bibliography{Biblio} 



\appendix

\section{Other intensity mapping predictions}
\label{sec:IM}

\begin{figure}
	\includegraphics[trim={0cm 3 0 0},clip,width=1.05\columnwidth]{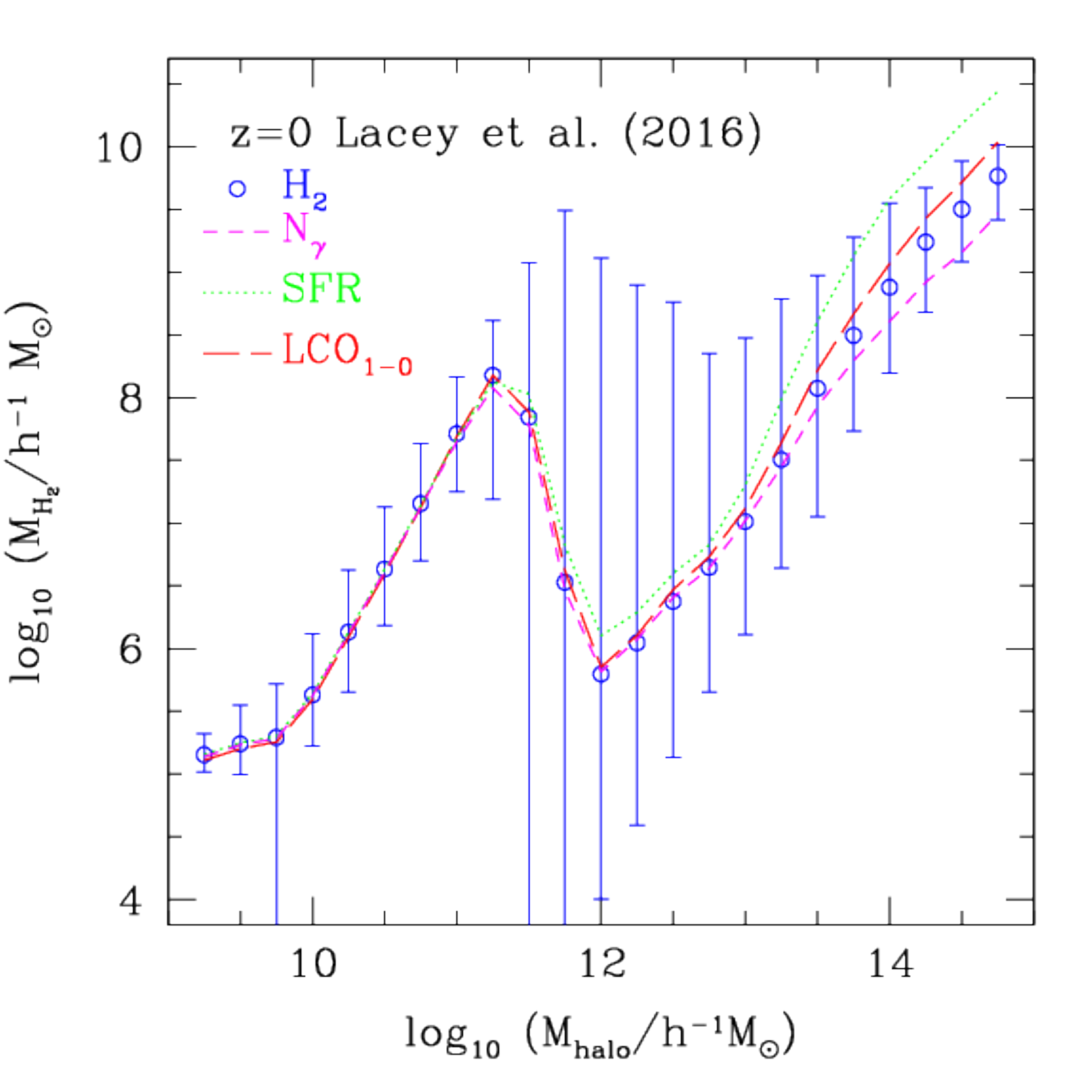}
    \caption{Various galaxy properties plotted against host halo mass. The y-axis shows the total $H_{2}$ gas mass as a function of halo mass; other quantities have had a single scaling factor applied to plot them on the same axis (the factors applied to the logarithm of the property are: the luminosity in the $1$-$0$ CO line luminosity measured in units of $10^{40} \,h^{-2} {\rm erg} \,{\rm s}^{-1}$, $11.57$, the star formation rate output in units of $h^{-1} {\rm M_{\odot}} {\rm yr}^{-1}$, $8.82$, the number of Lyman continuum photons expressed in $10^{40} {\rm s}^{-1} {\rm M_{\odot}}^{-1} $, $-4.15$). 
    }
    \label{fig:xmhalo}
\end{figure}

Here we consider other properties of galaxies that are relevant to the intensity mapping of various emission lines in the case of the recalibrated Lacey et al. model.  Fig.~\ref{fig:xmhalo} compares various model predictions for galaxy properties after applying a universal rescaling factor in each case, so that the quantity can be plotted on the same scale as the total H$_{2}$ mass of halos. The plot shows how the following properties vary with halo mass: (i) the total mass of molecular hydrogen, H$_{2}$, which is the fuel available for star formation, (ii) the star formation rate, (iii) the total number of Lyman continuum photons emitted per unit time, which is one of the factors driving the intensity of emission lines from ionized gas and (iv) the luminosity of the 1-0 transition in CO (see \citealt{Lagos:2012} for an explanation of the modelling of CO emission in from photon dominated regions {\tt GALFORM}). These quantities display a remarkably similar dependence on halo mass to one another, with a slight variation in the dependence on halo mass at high masses. This is expected, given that the in the model the star formation rate is proportional to the mass of molecular hydrogen. As we saw for the HI content of haloes, there is a pronounced break in the relation at the halo mass for which AGN feedback becomes important. There is also considerable scatter in how the properties scale with halo mass around this break.  

\section{Sensitivity of predictions to the treatment of gas cooling in satellite galaxies}
\label{sec:ram}

\begin{figure}
	\includegraphics[trim={0.7cm 1.7cm 0 0.4cm},clip,width=1.05\columnwidth]{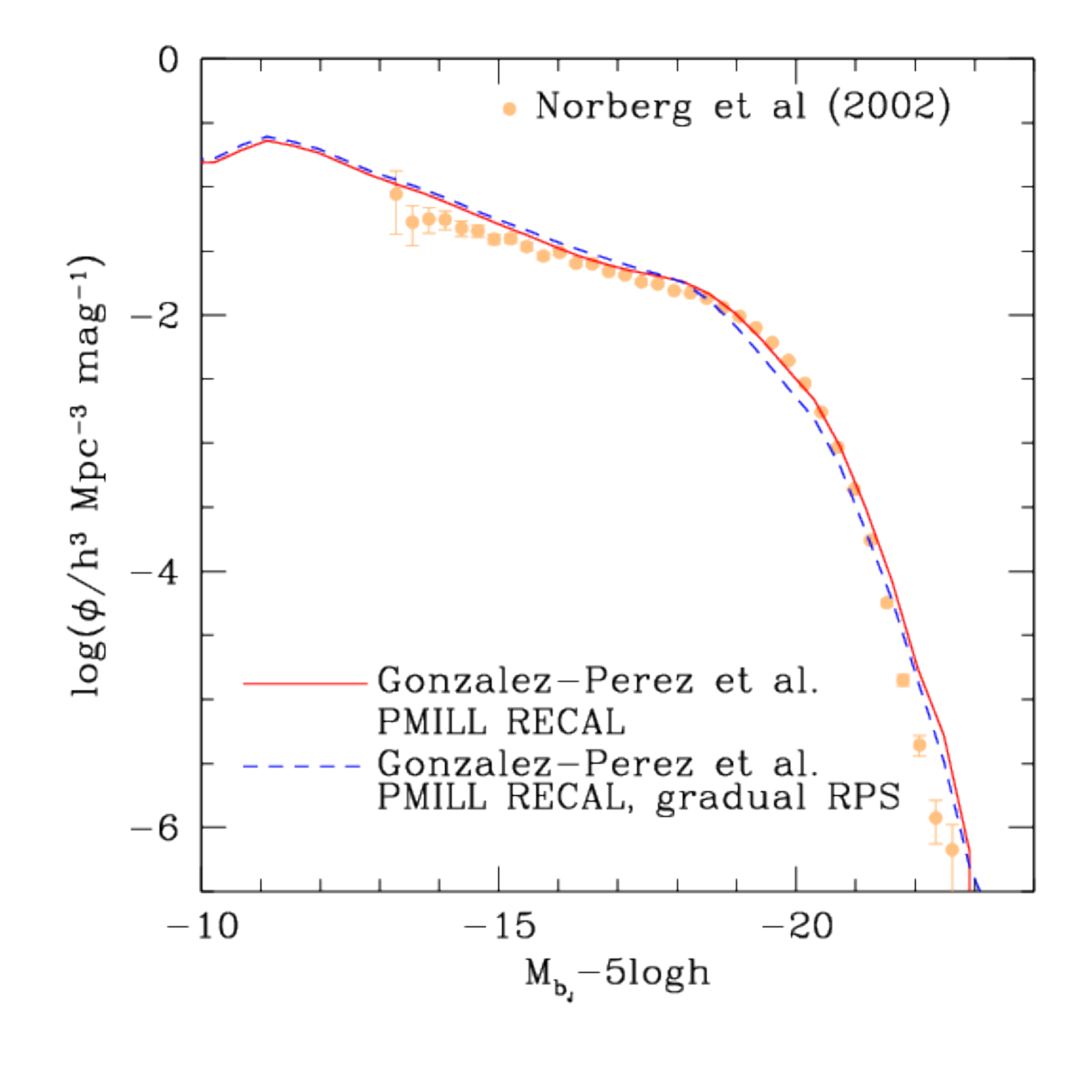}
    \caption{The effect of changing the gas cooling in satellites on the local galaxy luminosity function. The solid red line shows the version of the Gonzalez-Perez et~al. model recalibrated for the PMILL. In this model, the hot halo of a galaxy is assumed to be stripped instantaneously as soon as it becomes a satellite galaxy. The dashed blue line shows the model prediction for the luminosity function is this hot gas is stripped gradually, based on the ram pressure within the host dark matter halo.}
    \label{fig:lfBJ_grps}
\end{figure}

\begin{figure}
	\includegraphics[trim={0.7cm 1.7cm 0 0.4cm},clip,width=1.05\columnwidth]{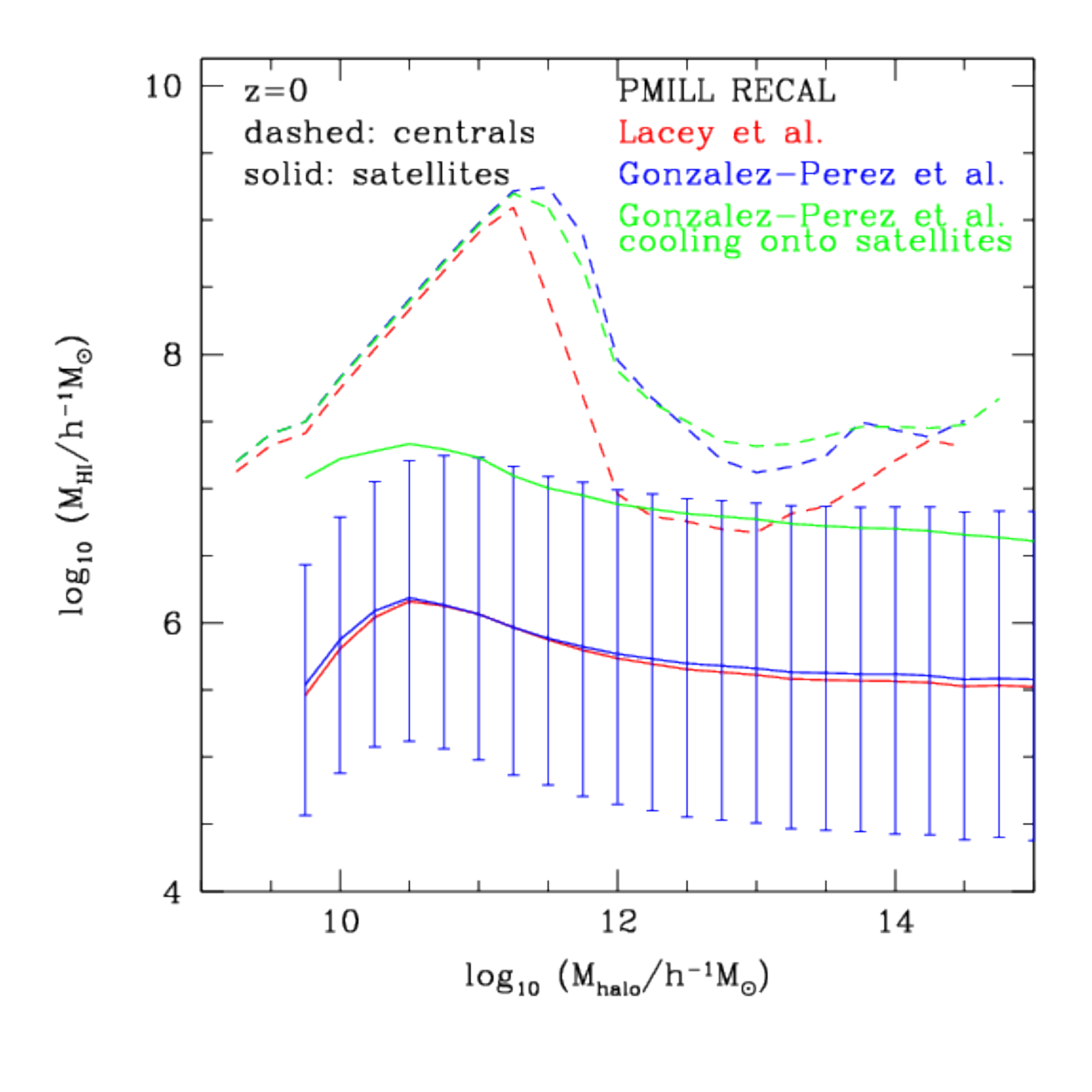}
    \caption{The (number weighted) median HI galaxy mass as a function of host halo mass, showing central galaxies (dashed) and {\em individual} satellite galaxies (solid). NB all curves in this plot are calculated for individual galaxies. The bars show the 10-90 percentile range of the distribution of satellite galaxy HI masses for the Gonzalez-Perez et~al. model. The different colours  show different models as indicated by the legend; note that the green curves show a variant of the Gonzalez-Perez et~al. model in which gradual ram pressure stripping of hot gas in satellites is invoked.}
    \label{fig:himhalo1}
\end{figure}

\begin{figure}
	\includegraphics[trim={0.7cm 1.5cm 0 0.4cm},clip,width=1.05\columnwidth]{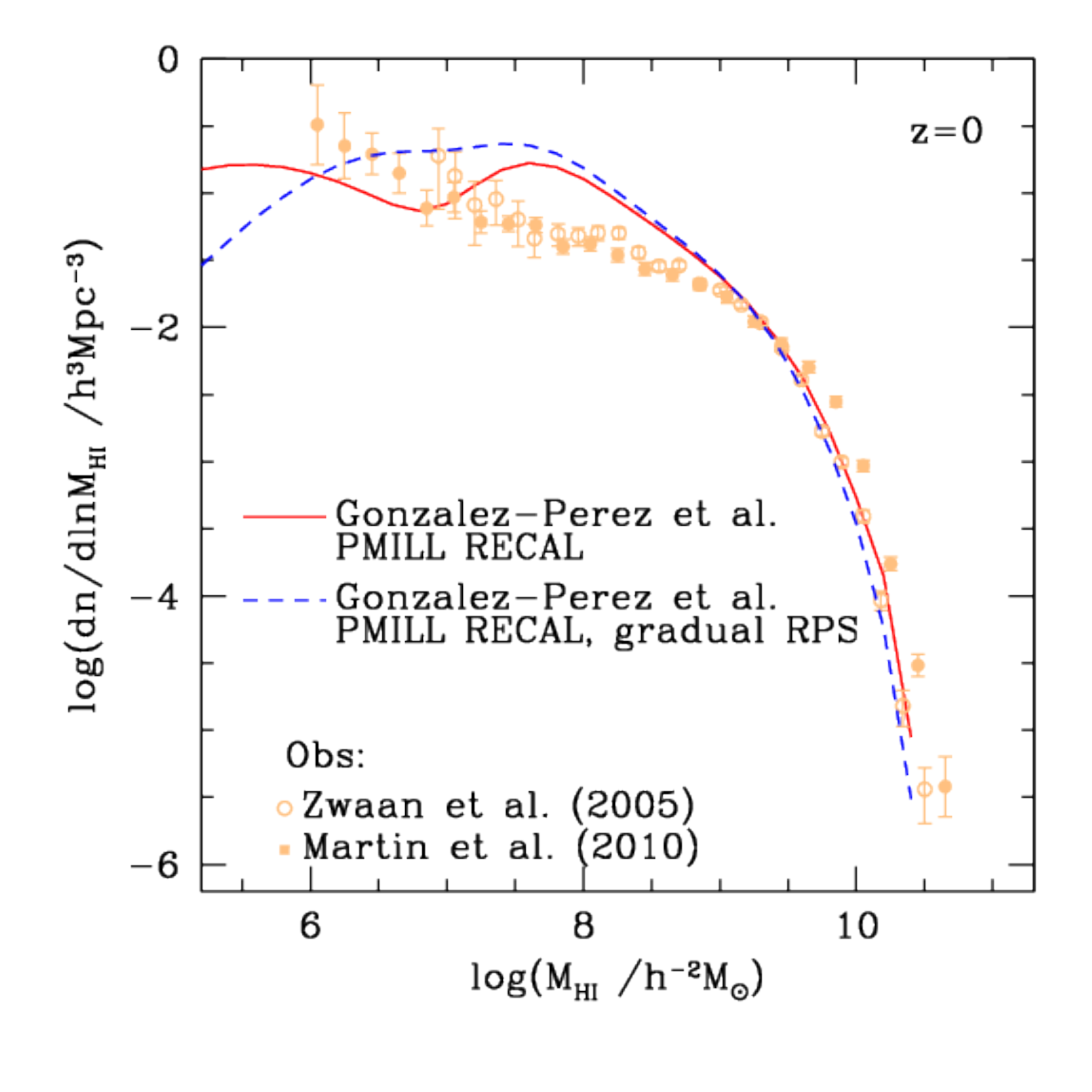}
    \caption{The influence of gradual ram pressure stripping on the $z=0$ HI galaxy mass function. The solid red line shows the recalibrated version of the Gonzalez-Perez et~al. model with instantaneous stripping of the hot gas halo of satellite galaxies. The blue dashed line shows the model predictions when gradual ram pressure stripping of the hot gas is adopted.}
    \label{fig:himf_grps}
\end{figure}

\begin{figure}
	\includegraphics[trim={0.7cm 1.5cm 0 0.4cm},clip,width=1.05\columnwidth]{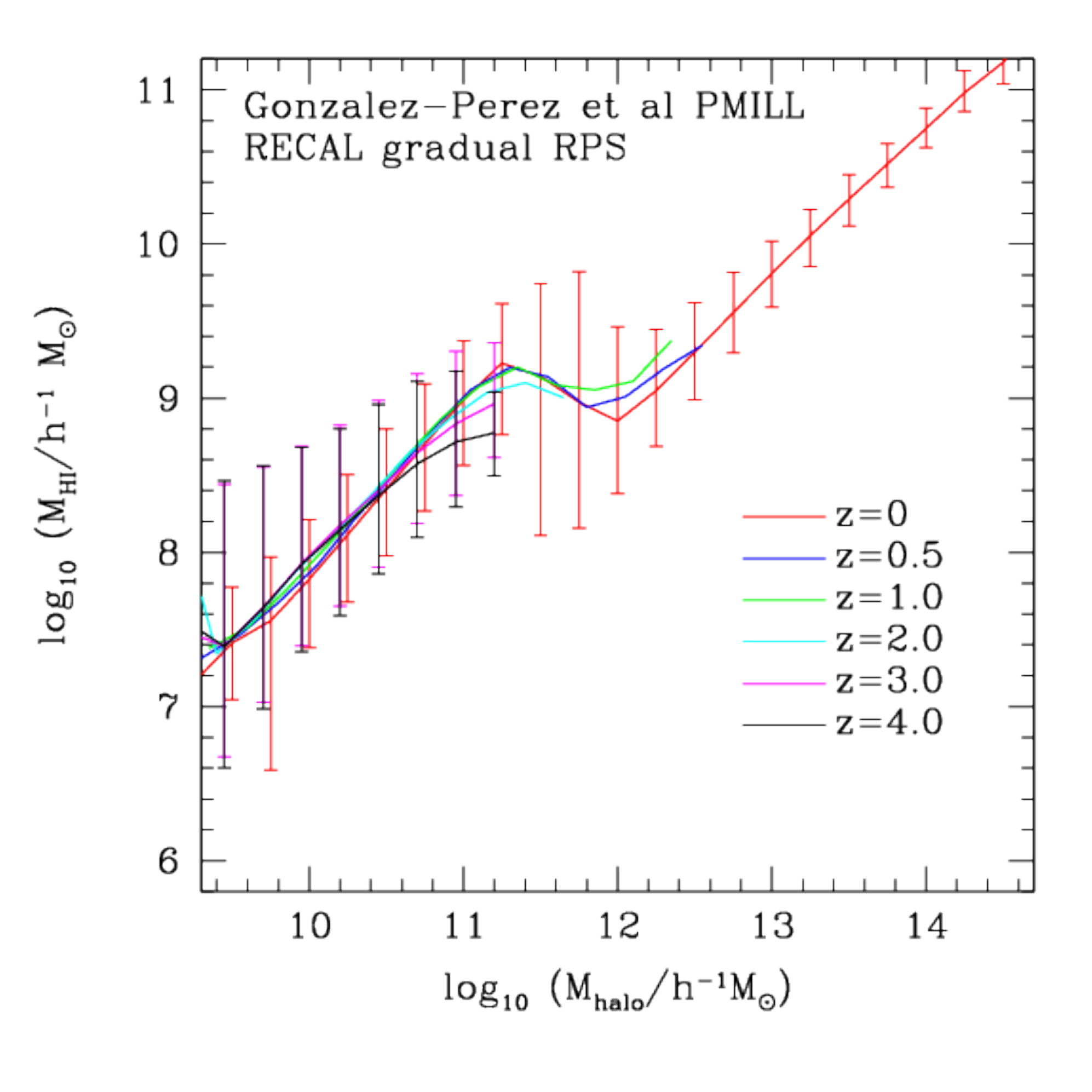}
    \caption{The HI mass $-$ halo mass relation in the recalibrated Gonzalez-Perez et~al model when gradual ram pressure stripping of the hot gas in satellite galaxies is adopted. The lines show the predictions at different redshifts as indicated by the key.}
    \label{fig:himhalo1_evol_grps}
\end{figure}

The assumption applied about gas cooling in satellite galaxies in the {\tt GALFORM} models discussed in the main part of this paper is that the hot gas halo of a satellite is instantly stripped away by the ram pressure of the hot gas in the main halo as soon as the galaxy becomes a satellite, and is added to the main hot gas halo. As a result, no gas cools onto satellite galaxies in these models. \cite{Font:2008} introduced a model with a gradual ram pressure stripping of the satellite gas, based on the hydrodynamic simulations of \cite{McCarthy:2008}.  A fraction of the cold gas that is reheated by supernova feedback, typically chosen to be 10\%, is also allowed to be stripped. This produces bluer, more gas-rich satellite galaxies, giving a better match to observational estimates of the fraction of passive galaxies at low stellar masses (see also \citealt{Lagos:2014a,Guo:2016,GP:2018}). 

Here we investigate the impact of adopting gradual ram pressure stripping of the hot gas in satellite galaxies in the recalibrated version of the Gonzalez Perez et~al. model used in the main paper. We do not change any of the other galaxy formation parameters. The impact on the local galaxy luminosity function is shown in Fig.~\ref{fig:lfBJ_grps}. The Gonzalez-Perez et~al. PMILL RECAL model with instantaneous stripping is shown by the solid red line. The variant with gradual stripping of the hot gas halos of satellite galaxies is shown by the blue dashed line. There is a small reduction ($\approx 25$\%) in the number of galaxies predicted around $L_*$ with gradual ram pressure stripping. This deficit could be reduced by adjusting other model parameters, such as those governing the timescale for gas heated by supernovae to be returned to the hot halo or the strength of AGN feedback (see \citealt{GP:2018}). We have not carried out this exercise here.

Fig.~\ref{fig:himhalo1} shows the effect of gradual ram pressure stripping on the mass of an {\it individual} satellite. With gradual ram pressure stripping, the median HI mass of satellites is over an order of magnitude higher than it is in the models with instantaneous ram pressure stripping of the hot gas in satellites (note that the precise values of the median HI masses are affected by the resolution of the simulation, which corresponds to a total mass in stars and cold gas of $10^{4} \, h^{-1} \, {\rm M_{\odot}}$, but this does not influence the comparison between models). The fact that the HI masses of satellites are so similar in the two RECAL models shows that the equilibrium reached in the gas content of galaxies is not sensitive to the precise values of the parameters used in the model, but instead depends on the generic form of the gas cooling, star formation and supernova feedback. The equilibrium reached in the case with gradual ram pressure stripping is fundamentally different, due to the different treatment of the reheated gas (see \citealt{Font:2008}). 

The HI mass function changes little at high masses with the variation in the treatment of gas cooling in satellites, as shown by Fig.~\ref{fig:himf_grps}. The shape of the mass function at low masses does depend on the treatment of gas cooling in satellites; however, the exact form of the model predictions in this mass regime is also affected by the resolution of the simulation (see Fig.~\ref{fig:himf_evol}).

The evolution of the HI mass $-$ halo mass relation in the model with gradual ram pressure stripping of the hot gas in satellites is shown in Fig.~\ref{fig:himhalo1_evol_grps}. As with the version of this model with instantaneous ram pressure stripping, there is little evolution on the relation until $z=4$. However, the depth of the break in this relation around a halo mass of $\approx 10^{11.5} h^{-1} {\rm M_{\odot}}$ is less pronounced in the model with gradual ram pressure stripping.


\bsp	
\label{lastpage}
\end{document}